\documentclass[10pt,journal,onecolumn]{IEEEtran}

\usepackage{amsmath,amsfonts}
\usepackage{algorithmic}
\usepackage{algorithm}
\usepackage{array}
\usepackage[caption=false,font=normalsize,labelfont=sf,textfont=sf]{subfig}
\usepackage{textcomp}
\usepackage{stfloats}
\usepackage{url}
\usepackage{verbatim}
\usepackage{graphicx}
\usepackage{cite}

\usepackage{mdframed}
\usepackage{titletoc}
\usepackage{etoolbox}
\usepackage{multirow}
\usepackage{amsthm}
\usepackage{mathpazo}

\usepackage{amssymb}
\usepackage{amsthm}
\usepackage{enumerate}
\usepackage{indentfirst}
\usepackage{latexsym}
\newtheorem{Theorem}{Theorem}

\newtheorem{Definition}[Theorem]{Definition}

\newtheorem{Remark}[Theorem]{Remark}
\usepackage{makecell}
\usepackage[numbers]{natbib}

\begin{document}

\title{A fully decentralized auditing approach for edge computing: A Game-Theoretic Perspective}


\author{Zahra~Seyedi,  Farhad~Rahmati,  Mohammad~Ali,   and  Ximeng~Liu
\thanks{Z. Seyedi, F. Rahmati, and  M. Ali  are  with the 
Department of Mathematics and Computer Science, Amirkabir University of 
Technology, Tehran, Iran. E-mails: 
zahraseyedi@aut.ac.ir, mali71@aut.ac.ir, 
frahmati@aut.ac.ir}
\thanks{X. Liu is with the Key Laboratory of Information Security of Network Systems, College of Computer and Data Science, Fuzhou University,
Fuzhou 350108, China. Email: snbnix@gmail.com.  
}}



\maketitle

\begin{abstract}
Edge storage presents a viable data storage alternative for application vendors (AV), offering benefits such as reduced bandwidth overhead and latency compared to cloud storage. However, data cached in edge computing systems is susceptible to intentional or accidental disturbances. This paper proposes a decentralized integrity auditing scheme to safeguard data integrity and counter the traditional reliance on centralized third-party auditors (TPA), which are unfit for distributed systems. Our novel approach employs edge servers (ES) as mutual auditors, eliminating the need for a centralized entity. This decentralization minimizes potential collusion with malicious auditors and biases in audit outcomes. Using a strategic game model, we demonstrate that ESs are more motivated to audit each other than TPAs. The auditing process is addressed as a Nash Equilibrium problem, assuring accurate integrity proof through incentives for ESs. Our scheme's security and performance are rigorously assessed, showing it is secure within the random oracle model, offers improved speed, and is cost-effective compared to existing methods. 

\end{abstract}


\section{Introduction}
The Internet of Things (IoT) has seeped into every aspect of our life. This has had a profound impact on human society. From communication and transportation to healthcare, the rapid growth of IoT technology has fueled the variety and intricacy of mobile and IoT software applications. However, the limited resources of IoT devices can hinder the performance of applications that need a high-energy supply and intensive processing power \cite{int1} \cite{int0} \cite{int-1}. Applications such as real-time video analysis, virtual/augmented reality \cite{int2}, and interactive gaming \cite{int3} require low latency, real-time processing, and a large amount of bandwidth, making cloud computing an unsuitable computing method to meet such requirements \cite{int4}. Therefore, online application vendors (AVs) need a solution that wins out over cloud computing in latency-sensitive situations. To overcome these challenges, AVs can turn to distributed edge computing (EC) as an extension of the cloud computing paradigm from centralized cloud to edge servers \cite{int5}.

\begin{figure}[!ht]
	\centering
	\includegraphics[scale=.065]{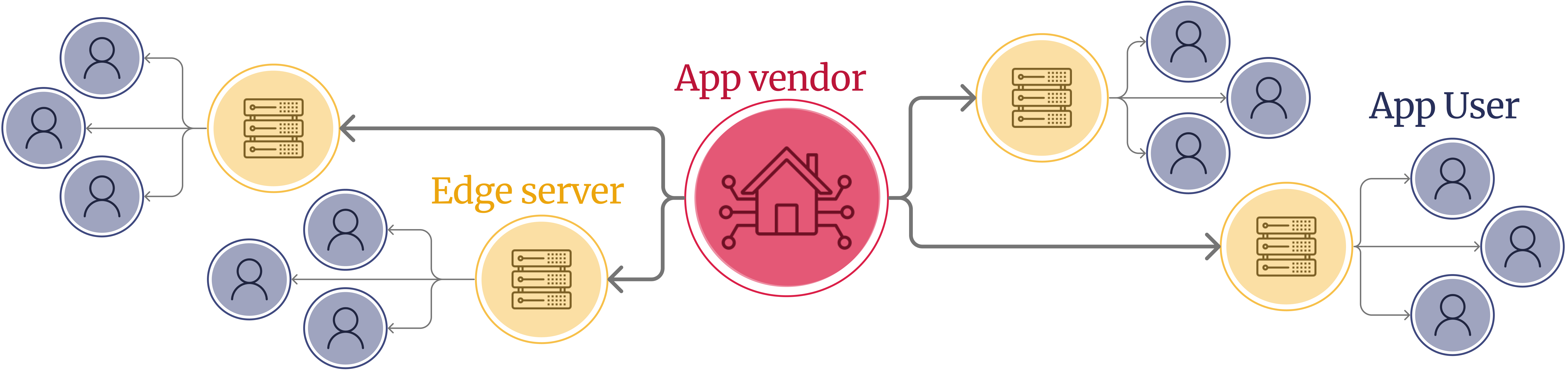}
	\caption{A caching system based on Edge Computing.}
	\label{sys model}
\end{figure}

EC is a paradigm that enables online applications to perform low latency and prompt processing by bringing computing resources near end-users \cite{int6} (Fig. \ref{sys model}). However, EC also introduces security challenges, as it involves caching data to multiple edge servers (ESs) that may be vulnerable to data corruption \cite{int7}. To ensure the integrity of their cached data, AVs need to perform data auditing, which is a complex and costly task in a large-scale and dynamic EC system \cite{int8}. A common solution is to delegate the data auditing to a third-party auditor (TPA), who can audit and authenticate the data on behalf of the AVs. However, this solution also poses some difficulties described below.

Hiring TPAs is expensive and requires a lot of computing power for data auditing in ESs. Also, AVs may pay a lot to hire them, but there are still challenges with timely auditing. TPAs may audit periodically or on demand, but these operations may not be fast or accurate. This can cause user problems if corrupted data is not audited and fixed quickly. Also, TPAs may be unreliable and dishonest, collaborating with ESs to fake or send false auditing results without proper verification. This can result in manipulated data in ESs, posing security risks \cite{int10}.

The main issue contributing to the aforementioned challenges is the centralized nature of TPAs, which contradicts the distributed nature of edge computing. While ESs are dispersed across various geographic locations, a centralized TPA assumes responsibility for verifying their data. To keep a distributed storage, it is better to use a distributed auditing method that matches this nature. When a system has centralized auditing, it becomes centralized, as its security depends on a few entities. Despite the appearance of decentralization, it is important to note that many schemes in this field, including those utilizing blockchain, tend to be centralized in practice, as clarified in Section \ref{rw}. Moreover, as the number of ESs grows, a centralized TPA may face challenges in meeting the prompt demands of AVs. Additionally, a centralized remote TPA needs a lot of bandwidth for data transfer, such as auditing and proof exchange between ESs and the TPA. This can be very costly \cite{int10}.

\begin{figure}[!ht]
	\centering
	\includegraphics[scale=.08]{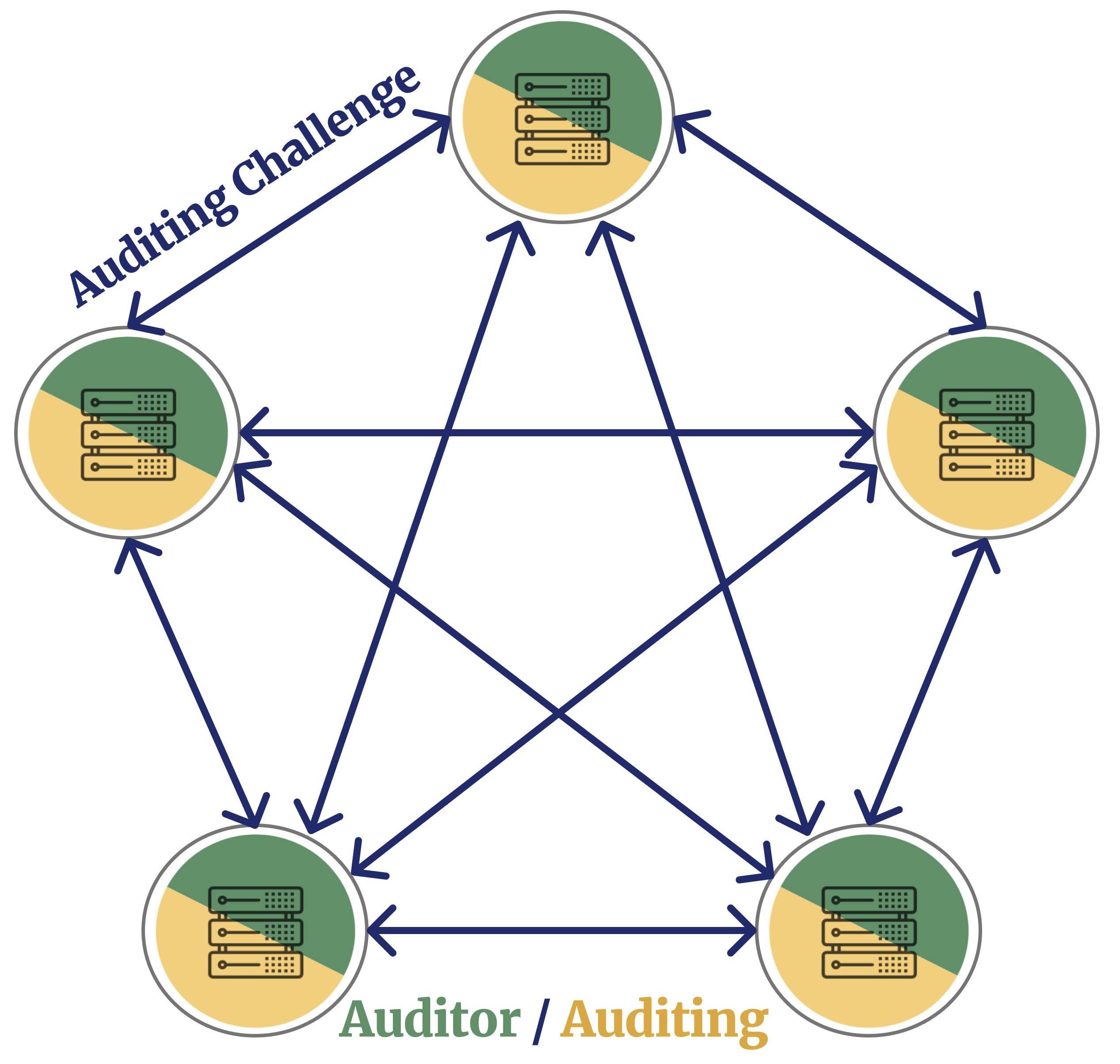}
	\caption{Fully distributed auditing architecture.}
	\label{pentagon}
\end{figure}

\subsection{Main Contributions}
To manage the discussed challenges, we propose a Fully Decentralized Integrity Auditing (FDIA) scheme in this paper. Our scheme eliminates the need for a centralized third-party auditor (TPA) and allows edge servers (ESs) to audit each other’s data integrity. We should note that the idea used in our remote auditing mechanism has been first introduced in \cite{O2RI}. However, \cite{O2RI} assumes a centralized architecture for the auditing mechanism, and in this work, we make it decentralized and distributed.

\begin{itemize}
	\item \textit{Distributed integrity auditing}: Our scheme replaces TPA with ESs, making the auditing process fully distributed. Our scheme is the first to achieve full distribution for both auditing and storage. This approach eliminates the need for AVs to hire a third party, as each ES audits the others (as in Fig. 2). Moreover, this scheme solves some of the mentioned challenges by increasing the auditors and their distribution. Also, decentralizing auditors and allowing multiple entities to audit helps remove the challenge related to auditor unreliability.

	\item \textit{Game-assisted auditing mechanism}: Using a strategic game and utility functions of entities, we show that ESs are more motivated than TPAs to audit each other, making them a more suitable choice for integrity verification. Furthermore, we employ this game theoric mechanism to scrutinize and incentivize ESs to provide truthful and accurate integrity proof.
	
	\item \textit{Efficient auditing process}: To protect the main task of serving app users, our FDIA scheme has low computational overhead for ESs during data auditing. Our approach is efficient and effective in terms of computational complexity and communication complexity of the data integrity inspection protocol, as well as storage overhead on auditor ESs. Furthermore, since ESs have limited computing power, our auditing process is optimized to reduce computing overhead for newly verified ESs during replicated audits of data integrity.
	
	\item \textit{Provable security}: 
	Our FDIA approach includes a system model, a threat model and security definitions. Additionally, we demonstrate the security of FDIA by proving its resilience against attacks according to the bilinear Diffie-Hellman (BDH) problem's hardness.
	
\end{itemize}

\subsection{Paper Organization}
This paper are structured in the following manner: related work and preliminaries are discussed in Sections \ref{rw} and \ref{pre}, respectively. Section \ref{pf} describes problem formulation. Section \ref{fdia} provides the details of the FDIA scheme. A game overview and security analysis of the FDIA are provided in Sections \ref{gta} and \ref{sa}, respectively. The performance evaluation and results are presented in Section \ref{per}. Finally, the paper concludes in Section \ref{cn}.

\section{Related Work}\label{rw}
Edge computing is a new distributed computing paradigm which plays a vital role in reducing latency and conserving bandwidth by employing computing and storage units toward the network's periphery. This technology enables AVs to hire ESs in different geographic locations, providing prompt services to their users \cite{rw9}. However, the development of this technology has led to a set of new challenges. Several solutions have been proposed to address some of these challenges, including computation offloading \cite{rw10}, edge server deployment \cite{rw12}, and edge user allocation \cite{rw14}. Meanwhile, issues related to edge data caching and data retrieval processes have received significant attention due to their numerous applications and great importance \cite{rw16}. Edge data caching systems need specific security requirements, especially data integrity and authentication. The cached data in ESs face threats and attacks \cite{int8}, \cite{int9}, so designing a secure and fair data integrity auditing scheme is challenging.

Data integrity auditing has been widely studied, and various auditing protocols have been developed, mainly provable data possession (PDP) and proofs of retrievability (PoR). PoR, initially proposed by Jules et al. \cite{rw21}, audits critical information like confidential data from sensitive units \cite{rw25}. This mechanism can detect and recover corrupted data. The efficiency of PoR depends on preprocessing tasks by the data owner prior to data upload, such as adding sentinels and correcting errors. However, PoR has limitations in updating and challenges due to the limited sentries. To this end, Shacham et al. \cite{rw26} created a new PoR method that enables public verification through multiple challenges, while preventing dynamic data corruption. During the auditing process, a considerable volume of data needs to be transferred from the server to local storage for verification and comparison. This leads to increased expenses in terms of communication, computational resources, storage capacity, and network bandwidth usage for those who own the data.

Ateniese et al., introduced a new approach to remote data integrity checking called provable data possession (PDP) using homomorphic tag technique \cite{rw18}. PDP is a probabilistic model that can efficiently verify the integrity of remote data without downloading the entire dataset. During the auditing process, probabilistic auditing greatly reduces computing burden and communication costs. However, the initial PDP scheme lacked support for dynamic data operations like remove and insert. To address this limitation, a full-dynamic PDP scheme are developed in \cite{rw19} using authenticated flip table technique, although this approach incurs considerable computational overhead. Shen et al. \cite{rw29} proposed a method that employs a doubly linked information table (DLIT) and a location array (LA) to manage batch data authentication and dynamic updates. Nonetheless, as the files become more intricate, retrieving data becomes challenging due to the linked list's characteristics. To mitigate this, Su et al. \cite{rw30} introduced a binary-ordered Merkle tree based system with a local valid root node to reduce the verification path of data nodes. Meanwhile, Hariharasitaraman et al. \cite{rw31} suggested using a position-aware Merkle hash tree (HMHT), which can determine the position of nodes in relation to their parent nodes within the tree. However, if there is an enormous amount of cached data, the PMHT structure will become complicated, leading to reduced audit efficiency.

To alleviate the performance issue faced by app vendors in the auditing process, usually, a TPA is considered to handle verification tasks. Zhu et al. \cite{rw32} designed a data integrity inspection scheme by utilizing a short signature algorithm, enabling privacy protection and public inspection with TPA involvement. Wang et al. \cite{rw33} designed a TPA-based public auditing scheme that effectively reduces computational overhead for data owners. Also, they presented the Merkle tree model for dynamic data processing, but audit efficiency may decline with increased data volume. 
Lu et al. \cite{rw34} presented a lightweight auditing scheme where the TPA performs block-tag production and integrity inspection on behalf of the data owner. However, these traditional schemes fail to address the security concerns associated with TPA utilization, which we want to tackle in this paper. 

Blockchain-assisted approaches provide solutions to the challenges of TPAs and data auditing across multiple servers. Current blockchain-based research focuses on applications in cloud storage, IoT, and smart cities. By replacing TPAs with blockchain, auditing can be delegated and permanently recorded effectively. Huang et al. \cite{rw35} designed a collaborative auditing framework using consensus nodes as TPAs to preserve remote data integrity. A TPA-free blockchain framework using random challenging numbers and Merkle trees is proposed in \cite{rw36} for decentralized ECS data integrity inspection. Wei et al. \cite{rw37} enable multi-tenant cooperation through smart contracts on the blockchain to ensure data authentication.

Despite blockchain-based approaches being explored as a solution for addressing TPA-related concerns, these schemes have not fully resolved the challenges and have introduced new concerns. It is known that the effectiveness of a blockchain network depends on the quality of its nodes, rendering it non-neutral \cite{rw38}. Moreover, if a single entity controls over 50\% of the nodes, the network becomes insecure. The storage requirements for multiple nodes present a challenge, with blockchain size expanding alongside transaction and node growth. As another challenge, the implementation of blockchain is costly, necessitating significant investments to incentivize nodes, maintain the technology, and cover associated expenses, potentially surpassing TPA expenses \cite{rw38} \cite{rw39}. 

After reviewing the related work above, we can see that traditional methods are not adequate to meet current auditing requirements. Although some scholars have attempted to incorporate blockchain technology into remote auditing systems, these approaches are flawed and pose numerous issues. Therefore, there is a need for a more comprehensive approach to enhance authentication security, minimize the computational burden of multiple authentications, and improve overall performance. This paper aims to address these concerns and provide novel contributions to the field. Table \ref{comparisonTableRW} compares FDIA with the described state-of-the-art methods. 

		\begin{table}[!ht]
		\begin{center}

			\caption{Comparison of various schemes functionalities}\label{comparisonTableRW}
			\scalebox{1}{
			\begin{tabular}{l|cccccccccc}
				\hline
				& \cite{int8} & \cite{rw2} & \cite{rw3}& \cite{rw4}& \cite{rw5}& \cite{rw6} & \cite{rw7} & \cite{rw8} & ours       \\ 
				\hline
				$\mathcal{F}_1$ & \checkmark & & & & \checkmark &  & & & \checkmark  \\ 
				\hline
				$\mathcal{F}_2$ & & \checkmark & & \checkmark & & \checkmark & \checkmark & \checkmark& \checkmark  \\ 
				\hline
				$\mathcal{F}_3$ &  & & & &\checkmark & & & \checkmark& \checkmark  \\ 
				\hline
				$\mathcal{F}_4$ &  & & \checkmark & & & & & \checkmark & \checkmark  \\ 
				\hline
				$\mathcal{F}_5$ & & & & & & & & &\checkmark  \\ 
				\hline
				$\mathcal{F}_6$ & & & & & & & & &\checkmark  \\ 
				\hline
				\hline
				\multicolumn{10}{l}{
				\begin{footnotesize}
				Notes. 
				$\mathcal{F}_1$: Distributed caching system,
				\end{footnotesize}
				}\\
				\multicolumn{10}{l}{
				\begin{footnotesize}
				$\mathcal{F}_2$: Low communication overhead of data auditing process,
				\end{footnotesize}
				}\\
				\multicolumn{10}{l}{
				\begin{footnotesize}				
				$\mathcal{F}_3$: Lightweight challenge generation process, 
				\end{footnotesize}
				}\\
				\multicolumn{10}{l}{
				\begin{footnotesize}				
				$\mathcal{F}_4$: Distributed Auditing system, 
				\end{footnotesize}
				}\\
				\multicolumn{10}{l}{
				\begin{footnotesize}				
				$\mathcal{F}_5$: Removal of the third party, 
				\end{footnotesize}
				}\\
				\multicolumn{10}{l}{
				\begin{footnotesize}
				$\mathcal{F}_6$: ES as auditor.
				\end{footnotesize}
				}
			\end{tabular}
			}
			\end{center}
		\end{table}
\section{Preliminaries}\label{pre}
In this section, we briefly state some of the prerequisites needed in the rest of this article.

\subsection{Bilinear map}\label{Bilinear map}
Suppose $G_1$ and $G_2$ are two cyclic groups of a prime order $p$. A function $e : G_1 \times G_1 \rightarrow G_2$ is considered bilinear if it satisfies the following three conditions:

\begin{itemize}
	\item[-]
	Bilinearity: For each $\alpha \in G_1$ and two members $x, y $ of $\mathbb{Z}_p$, we have $\sigma 
	(\alpha^x , \alpha^y) = e(\alpha^x , \alpha^y) = e (\alpha, \alpha)^{xy} $;
	
	\item[-]
	Non-degeneracy: There is an element $\alpha \in G_1$ such that $e (\alpha^x , \alpha^y) \neq 1$; 
	
	\item[-] 
	Computability: For any $\alpha , \beta \in G_1$, there exists an efficient algorithm to compute $e (\alpha , \beta)$.
	
\end{itemize}

\subsection{Bilinear Diffie-Hellman (BDH) problem}\label{DBDH}
Assume that $\mathcal{G}$ is a probabilistic polynomial-time (PPT) algorithm. It takes a security parameter $n$ as input and produces a tuple $(p, G_1, G_2, e)$, which is the same as Section \ref{Bilinear map}. The decisional bilinear Diffie-Hellman (DBDH) assumption is considered to hold for $\mathcal{G}$ if, for any tuple $(n, p, G_1, G_2, e, \alpha, \alpha^x, \alpha^y , \alpha^z , e(\alpha, \alpha)^w)$, in which $n$ is the security parameter, $(p, G_1, G_2, e)$ is the output of $\mathcal{G}(1^n)$, and $\alpha \in G_1$ and $x,y,z \in \mathbb{Z}_p$ are chosen uniformly at random, the advantage of any PPT distinguisher $\mathcal{D}$ in distinguishing the case of $w$ being either equal to $xyz$ or a random element of $\mathbb{Z}_p$ is negligible in $n$.

\subsection{ID-based Aggregate Signature}
Using an aggregate signature scheme \cite{pre1} a single verifier can combine multiple signatures created by different users on different messages, thereby reducing the computational overhead required to verify each signature individually. An ID-based aggregate signature (IDAS) scheme $\Pi$ typically includes five PPT algorithms $(\textbf{Setup}_{sign}, \textbf{KeyGen}_{sign}, \textbf{Sign}, \textbf{Agg}, \textbf{Vrfy})$:

\begin{itemize}
	\item $\textbf{Setup}_{sign}(1^n)$: This algorithm generates public parameters $pk_{sign}$ and the master secret key $msk_{sign}$ using security parameter $n$.
	\item $\textbf{KeyGen}_{sign}(pk_{sign}, msk_{sign}, id)$: Given the public parameters $pk_{sign}$, the master secret key $msk_{sign}$, and an identifier $id$, it returns a secret key $sk_{id}$.
	\item $\textbf{Sign}(pk_{sign}, id, sk_{id}, M)$: The algorithm takes four inputs: a message $M$, public parameters $pk_{sign}$, identifier $id$, and secret key $sk_{id}$. It returns a signature $\sigma$.
	\item $\textbf{Agg} (pk_{sign}, \lbrace \sigma_j \rbrace_{j \in J} )$: Given the public parameters $pk_{sign}$ and a set of signatures $\lbrace \sigma_j \rbrace_{j \in J}$, this algorithm produces an aggregate signature $Agg(\lbrace \sigma_j \rbrace_{j \in I})$.
	
	\item $\textbf{Vrfy}(pk_{sign}, Agg(\lbrace \sigma_i \rbrace_{i \in I}), \lbrace id_j \rbrace_{j \in J} , \lbrace M_j \rbrace_{j \in J})$: 	The algorithm takes public parameters $pk_{sign}$, an aggregate signature $Agg(\lbrace \sigma_i \rbrace_{i \in I})$, identifiers $\lbrace id_j \rbrace_{j \in J}$, and message $\lbrace M_j \rbrace_{j \in J}$ as input and outputs a bit $b \in \lbrace 0, 1\rbrace$ such that $b = 1$ represents valid signature and $b = 0$ represents invalid signature.	
	
\end{itemize}

\subsection{Strategic Game}
A tuple $(P, \lbrace S_i \rbrace_{i \in P}, \lbrace f_i \rbrace_{i \in P}) $ is a strategic game consisting of a finite set of players $P$, each with a non-empty set of actions $S_i$ available to them and a payoff function $f_i$ on $S = \prod_{i \in P} S_i$ where $i \in P$.

We utilize $s_i \in S_i$ to represent player $i$'s strategy. A set of strategies for all players is given by $s = \lbrace s_i \rbrace_{i \in P}$ and is called a strategy profile. A set of strategies for all players exclude the $i$-th one is represented by $s_{-i} = (s_1 , s_2, ..., s_{i-1}, s_{i+1}, ..., s_p) \in S_{-i}$.

A \textit{Nash equilibrium} (NE) refers to a strategy profile $S^{*}$ where no player has the ability to enhance their own payoff by unilaterally changing their chosen strategy. To put it formally, the strategy profile $S^{*} = (s^{*}_i,s^{*}_{-i})$ is a Nash Equilibrium if for each $s_i \in S_i$ and $ i \in P$, $f_i(s^{*}_i,s^{*}_{-i}) \geqslant f_i(s^{*}_i,s_{-i})$.


\section{Problem formulation} \label{pf}
The initial part of this section introduces the system model of our FDIA scheme. Following that, the security model of the proposed scheme is explained.

\subsection{System Model}
\begin{figure}[h!]
\centering
	\begin{tabular}{c}
		\includegraphics[width=2.4in]{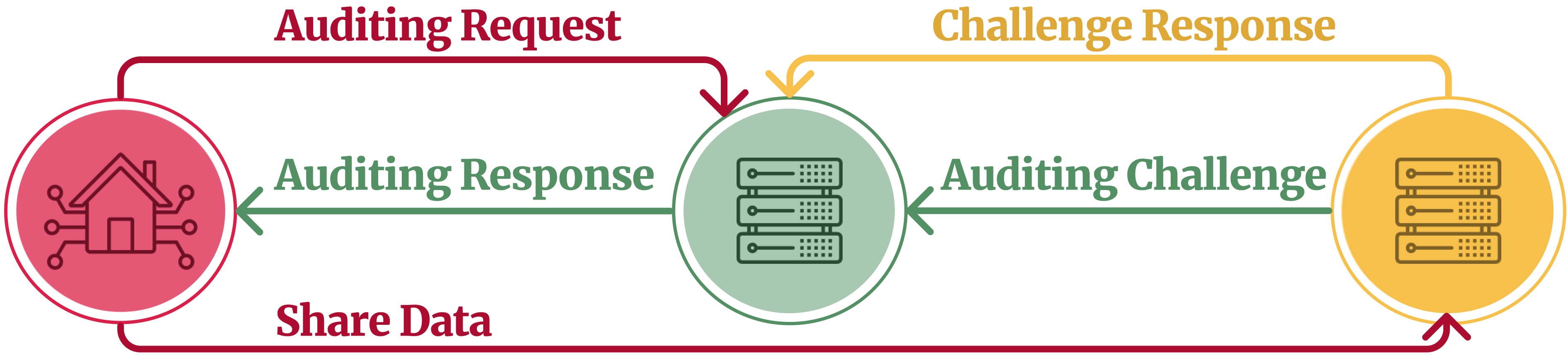} \\
		\scalebox{.8}{(a). Necessary auditing request. } \\
		\includegraphics[width=3.2in]{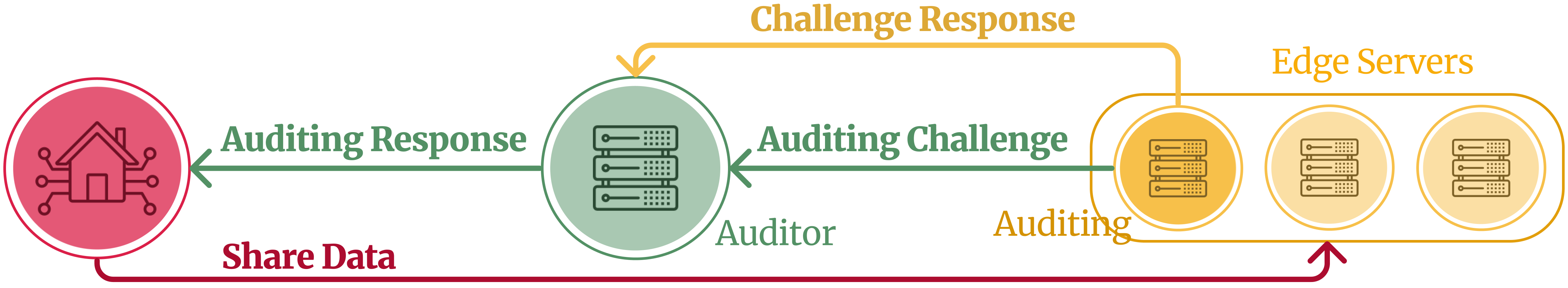} \\
		 \scalebox{.8}{(b). Randomly auditing.}\\
	\end{tabular}
	\caption{System models.}\label{fig_sim}
\end{figure}

There are two types of entities in our system: an app vendor (AV) and several edge servers (ES). ESs are employed by the AV to store their data in various location areas and serve app users faster (as shown in Fig. \ref{sys model}). Furthermore, ESs have an additional role in our system.  Each ES can audit cached data in other ESs either regularly, randomly or upon AV's request. In this EC-based system, the AV outsources auditing to ESs in two ways: 1. Regular or random inspection 2. Necessary and in-the-moment inspection. Each ES can be responsible for regularly or randomly auditing cached data of some other ESs on behalf of the AV (as illustrated in Fig. \ref{fig_sim} (b)). Alternatively, an ES can perform necessary audit requests received from the AV (as depicted in Fig. \ref{fig_sim}(a)).

\noindent
Proposed FDIA system involves eight algorithms namely \textbf{Setup}, \textbf{AV.keyGen},  \textbf{ES.keyGen}, \textbf{TagGen}, \textbf{Challenge}, \textbf{ProofGen}, \textbf{UpdateProof}, and \textbf{ProofCheck}.

\begin{itemize}
	\item $\textbf{Setup}(1^n)$: This is a probabilistic algorithm that is operated by the AV. Given a security parameter $n$, it returns the system parameters $params$, the master secret key $msk$, and its own secret key $sk_{AV}$.

	\item $\textbf{AV.keyGen}(params , msk , id_{AV})$: The AV runs this probabilistic algorithm.  It takes as input the system parameters $params$, the master secret key $msk$, and the AV's identity $ID_{AV}$, outputs the secret key $sk_{AV}$.
	\item $\textbf{ES.keyGen}(params , msk , id_{ES})$: The AV executes this probabilistic algorithm. Given the system parameters $params$, an ES's identity $id_{ES}$, and the master secret key $msk$, it returns the secret key $sk_{ES}$.
	\item $\textbf{TagGen}(params , F , sk_{AV})$: The AV operates this probabilistic algorithm. It takes as input the system parameters $param$, the AV's secret key $sk_{AV}$, and a file $F$ to be stored. For each file block $m_i$, it returns  tags $t = (t_1,..., t_n)$, that will be cached on ESs' memory alongside the file $F$.
	\item $\textbf{Challenge}(params, m , id_{ES}, sk_{ES})$: An auditor ES run this randomized algorithm. Given the system parameters $params$, the number of blocks $m$, and its identifier and secret key $id_{ES}$ and $sk_{ES}$ as inputs, it returns a challenge $ch_{F}$ on behalf of the AV.
	\item $\textbf{ProofGen}(params , ch_{F}, t, proofs_{F})$: This is a probabilistic algorithm that is operated by the auditing ES with inputs including the system parameters $params$, the challenge $ch_{F}$, the tag $t$, and the proofs of this file $proofs_{F}$ which it cached in its memory, to output a data integrity proof $P_{F}$ of the challenged blocks and parameters $R_{update}$ and $R_{ch}$ to update its state and proof check, respectively.

	\item $\textbf{UpdateProof}(I , I', proofs_{F}, R_{update})$: An ES runs this algorithm. It takes as inputs the proofs state $proofs_{F}$, and parameters related to updating the proofs state $R_{update}$, updates the proofs state $P_{F}$.				
	\item $\textbf{ProofCheck}(params, ch_{F}, P_{F}, R_{ch}, name_{F})$:
 	This deterministic algorithm is executed by the auditor ES. Given as inputs the system parameters $params$, the challenge $ch_{F}$, the file name $name_{F}$, a proof $P_{F}$, and the proven blocks history $R_{ch}$, it outputs either 1 or 0,  indicating whether the file $F$ remains unchanged.
\end{itemize}

\subsection{Security Model}
The main security concerns of our proposed system are edge data corruption and forging data integrity-proof threats. We briefly describe them in the following:
\begin{itemize}
	\item[-]
	Edge data corruption: Cached data in ESs are subject to all kinds of intentional or accidental attacks and may be tampered with or corrupted. Thereby, ESs are not reliable and valid in this respect and are not assumed to be trustworthy. 
	\item [-]
	Forging data integrity proof: Additionally, ESs may hide the destruction or loss of data from the AV to gain more benefits or avoid paying compensation. Indeed, despite the loss or destruction of cached data, ESs may attempt to provide a proof of integrity to the auditor that deceives him/her about the accuracy of this data.
\end{itemize}

In order to address the aforementioned threats, our FDIA scheme should encompass the following features:
\begin{itemize}
	\item 
	\textit{Unforgeability}: This requirement mandates that the ESs, whose data has been corrupted, should not possess the capability to provide a valid integrity proof. In other words, the system should prevent any falsification of evidence by compromised ESs.
	\item 
	\textit{High efficiency}: Our proposed method should achieve a high level of efficiency for both the AV and ESs. The AV should be able to rapidly tag the data files and remotely audit the cached data files with minimal computational expenses after caching them in ESs. Additionally, ESs should be capable of verifying the integrity of the data files with low computation and communication overhead.
	\item  
	\textit{Effectiveness}: 3. Effectiveness: Our proposed approach should enable the AV to easily detect any corruption in data files within the edge data.
\end{itemize}

\begin{figure*}[ht!]
	\centering
	\includegraphics[scale=.16]{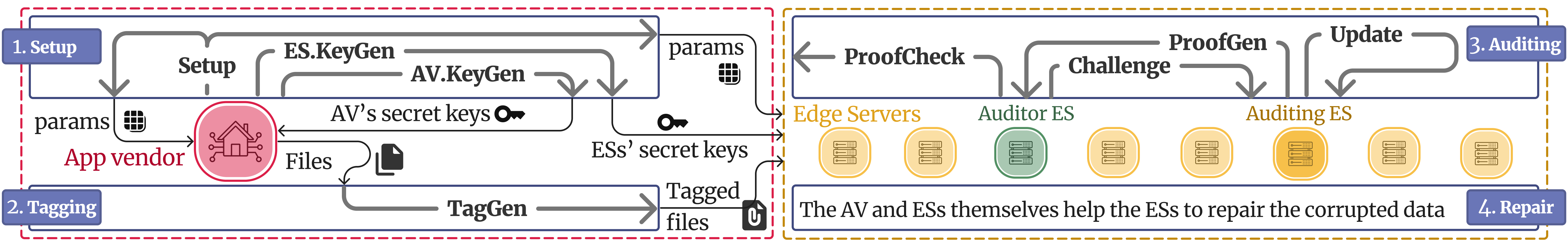}
	\caption{Scheme overview}
	\label{overview}
\end{figure*}
\section{FDIA scheme}\label{fdia}
\subsection{Scheme overview}
In this section, we present our proposed FDIA scheme in detail. Fig. \ref{algorithms} describes the algorithms employed in FDIA. Our approach comprises four fundamental phases: \textbf{Setup}, \textbf{Tagging}, \textbf{Auditing} and \textbf{Repair} which are briefly described below (see Fig. \ref{overview}).

\begin{figure*}
	\begin{mdframed}
	\begin{small}	
	$\textbf{Setup}(1^n)$:
	It takes a security parameter $n$, executes $(n, p, G, G', e) \leftarrow \mathcal{G}(1^n)$ and selects $g \in G$ and $z \in \mathbb{Z}_p$	uniformly at random. Then it sets $h = g^z$ and selects an identifier $id^{*}$. Also, it selects two hash functions $H_1, H_2 : \lbrace 0, 1\rbrace ^{*} \rightarrow G$ and a hash function $H_3 : G' \rightarrow \lbrace 0, 1\rbrace ^{*}$, a pseudorandom function $PRF : \mathbb{Z}_p \times \mathbb{Z}_p \rightarrow \mathbb{Z}_p$, and a secure IDAS scheme $\Pi = (\textbf{Setup}_{sign}, \textbf{KeyGen}_{sign}, \textbf{Sign}, \textbf{Agg}, \textbf{Vrfy})$. Then, by running $\textbf{Setup}_{sign}(1^n)$ it generates $(pk_{sign} , msk_{sign})$. Moreover it considers $m \in \mathbb{N}$ as the number of blocks in each data file. Finally, this algorithm returns $(msk , params)$ where $msk = (z, msk_{sign})$ and $params = (n, p, g, G, G', e, h, id^{*}, H_1, H_2, H_3, m, \Pi, pk_{sign})$.\\
				
	$\textbf{AV.KeyGen}(params , msk , id)$:
	Given the master secret key $msk = (z, msk_{sign})$ and the identity $id$, it runs $\textbf{KeyGen}_{sign}(pk_{sign}, \\ msk_{sign}, id)$ to generate a signature key $sk_{sign}$ and computes the private key $s = H_{1}(id)^z$. Finally, it outputs $(sk_{id}, s)$.\\
				
	$\textbf{ES.KeyGen}(params , msk , id)$:
	Given the master secret key $msk = (z, msk_{sign})$ and the identity $id$, this algorithm runs $\textbf{KeyGen}_{sign}(pk_{sign}, msk_{sign}, id)$ to output a signature key $sk_{id}$. \\ 
				
	$\textbf{TagGen}(params , F , sk_{id})$: 
	On input a file $F$ named $name_F$, it divides $F$ into $m$ blocks $f_1, ... , f_m$, where $f_i \in \mathbb{Z}_p$ and then chooses a random $r \in \mathbb{Z}_p$ and calculates $h' = g^r$ and $h" = h^r$. For each block $f_i$, it computes the tag $t_i = s^{f_i}H_{2}(name_F|| h"||i)^r$. Moreover it executes $ \sigma_{F} \leftarrow \textbf{Sign}(pk_{sign} , sk_{id}, (h'||name_F))$.	Then, it outputs the tuple $F_{t} = (h', t = \lbrace t_i \rbrace_{i = 1}^{m}, name_F, \sigma_{F})$.\\
				
	$\textbf{Challenge}(params, m, id, sk_{id})$: 
	It selects a random $k$-element subset $I$ of the set $[1, m]$. Also, it picks $k_{PRF} \in \mathbb{Z}_p$ and executes $ \sigma_{k_{PRF}} \leftarrow \textbf{Sign}(pk_{sign} , id, sk_{id}, k_{PRF})$. To produce a challenge, it selects a random $\lambda \in \mathbb{Z}_p$ and computes $\Lambda = e(H_1(id), h)$ and $\alpha = g^\lambda$, $\beta = \Lambda ^{\lambda}$. Finally it sends the challenge $ch_{F} = (\alpha, \beta, I, k_{PRF}, \sigma_{k_{PRF}}, id)$ to the auditing ES.\\
				
	$\textbf{ProofGen}(params, ch_{F}, t, proofs_{F})$:	
	Upon receiving $ch_{F} = (\alpha, \beta, I, k_{PRF}, \sigma_{k_{PRF}}, id)$, this algorithm initially calculates $\Lambda = e(H_1(id), h)$ and checks if $e(\alpha , \Lambda) = e(\beta , g)$. If not, the auditing aborts. Otherwise, according to $proofs = \lbrace ( I^{(j)} , \, k_{PRF}^{(j)}, \, \sigma_{k_{PRF}}^{(j)}, \, id^{(j)}, \, \lbrace \phi_{i} , \mu_{i} \rbrace_{i \in I^{(j)}}) \rbrace_{j = 1}^{l}$, it puts $ I'^{(j)} = I \cap I^{(j)}$ for any $j \in L = \lbrace 1, ..., l \rbrace$ and calculates $\mu'_{i} = f_{k_{PRF}}(i) f_i$ and $\phi'_{i} = t_{i}^{f_{k_{PRF}}(i)}$ for any $i \in I \setminus \cup_{j \in L} I'^{(j)}$. Then it computes $\mu = \sum_{j\in L}\sum_{i \in I'^{(j)}} \mu_{i} + \sum_{i \in I \setminus \cup_{j \in L} I'^{(j)}} \mu'_{i}$, $\phi = \prod_{j \in L}\prod_{i \in I'^{(j)}}\phi_{i} . \prod_{i \in I \setminus \cup_{j \in L} I'^{(j)}}\phi'_{i}$ and $m' = e(\phi, \alpha) . \beta^{-\mu}$ and returns $(P_{F}, R_{ch} , R_{update})$ where $ P_{F} = (m', h', \sigma_{F} , \lbrace I'^{(j)} \rbrace_{j \in L})$, $R_{ch} = \lbrace (I'^{(j)} , k_{PRF}^{(j)}, \sigma_{k_{PRF}}^{(j)}, id^{(j)}) \rbrace_{j \in L}$ and $R_{update} = (I \setminus \cup_{j \in L} I'^{(j)}, k_{PRF}, \sigma_{k_{PRF}}, id, \lbrace \mu'_{i} , \phi'_{i} \rbrace_{i \in I \setminus \cup_{j \in L} I'^{(j)}})$.\\
				
	$\textbf{UpdateProof}(I, I', proofs_{F}, R_{update})$: This algorithm takes $R_{update} = (I \setminus \cup_{j \in L} I'^{(j)}, k_{PRF}, \sigma_{k_{PRF}}, id, \lbrace \mu'_{i} , \phi'_{i} \rbrace_{i \in I \setminus \cup_{j \in L} I'^{(j)}})$ and add it to $proofs_{F} = \lbrace ( I^{(j)} , k_{PRF}^{(j)}, \sigma_{k_{PRF}}^{(j)}, id^{(j)}, \lbrace \phi_{i} , \mu_{i} \rbrace_{i \in I^{(j)}}) \rbrace_{j \in L}$.  \\
				
	 $\textbf{ProofCheck}(params, ch_{F}, P_{F}, R_{ch}, name_F)$: On input $m'$, it runs $\textbf{vrfy}_{sign}(pk_{sign}, id_{AV}, h'||name_F, \sigma_{F})$ to check if $\sigma_{F}$ is a valid identity-based signature on the message $h'||name_F$. If not, the proof is invalid. Otherwise, it runs  $\textbf{Agg} (pk_{sign}, \lbrace \sigma_{k_{PRF}}^{(j)} \rbrace_{j \in L} )$ to generate an aggregate signature and then runs $\textbf{vrfy}_{sign}(pk_{sign}, \lbrace id^{(j)} \rbrace_{j \in L}, \lbrace k_{PRF}^{(j)} \rbrace_{j \in L} , \lbrace \sigma_{k_{PRF}}^{(j)} \rbrace_{j \in L})$ to check if $\lbrace \sigma_{k_{PRF}}^{(j)} \rbrace_{j \in L}$ are valid signatures on keys $\lbrace k_{PRF}^{(j)} \rbrace_{j \in L}$. If not, the proof is invalid. Otherwise, it checks whether the following equality holds; if it does, the proof is accepted, and if not, it is deemed invalid.
	
\begin{footnotesize}			
\begin{align*}
H_{3}(m') = H_{3}(\prod_{j \in L}\prod_{i \in I'^{(j)}} e(H_2(name_F||h"||i)^{f_{k_{PRF}^{(j)}}(i)} , h'^{\lambda}). \prod_{i \in I\setminus \cup_{i \in L} I'^{(j)}} e(H_2(name_F||h"||i)^{f_{k_{PRF}}(i)} , h'^{\lambda})).
\end{align*}
\end{footnotesize}
\end{small}
\end{mdframed}
\caption{The concrete construction of our FDIA.}
\label{algorithms}
\end{figure*}

\begin{itemize}
	\item \textbf{Setup}: The AV initiates public parameters and its own secret keys that will be utilized in the subsequent phases, by running the algorithm $(msk_{AV}, params) \leftarrow \textbf{Setup}(1^n)$. It keeps $msk_{AV}$ completely confidential and makes $params$ publicly available. Then, it runs two key generation algorithms, $(sk_{AV}, s) \leftarrow \textbf{AV.KeyGen}(params , msk , id_{AV})$ and $sk_{ES} \leftarrow \textbf{ES.KeyGen}(params , msk , id_{ES})$ to generate ESs' and the AV's secret keys.
	
	\item \textbf{Tagging}: In this phase, the AV tags the data file to be shared with users of the system using the algorithm $F_{t} \leftarrow \textbf{TagGen}(params , F , sk_{id})$. Then, it sends the file $F$ along with its associated tag $F_{t}$, to the ESs  that wants to save the file. 
	
	\item \textbf{Auditing}: An ES as an auditor, $Ar$, checks the integrity of replicas cached in ESs by performing a challenge-response protocol. Given a data file $F$ cached in the auditing edge server $Ag$, the $Ar$ checks the integrity of $F$ without having $F$. It first executes $ch_{F} \leftarrow \textbf{Challenge}(params, m, id_{Ag}, sk_{Ag})$ to generate a challenge. It returns $ch_F$ to $Ag$. When $Ag$ receives the challenge, it calls the file $F$ and the tag $F_{t}$ associated with $F$. Then, it runs $(P_{F}, R_{ch} , R_{update}) \leftarrow \textbf{ProofGen}(params, ch_{F}, t, proofs_{F})$. $Ag$ update its proofs' state for file $F$ using the algorithm $\textbf{UpdateProof}(I, I', proofs_{F}, R_{update})$, and gives the proof $P_{F}$ along with $R_{ch}$ to the $Ar$. By running $\textbf{ProofCheck}(params, ch_{F}, P_{F}, R_{ch}, name_F)$, the $Ar$ checks the integrity proof.
	
	\item \textbf{Repair}: If an auditor ES, $Ar$ detects that cached file in auditing ES, $Ag$, has been corrupted, it refers the $Ag$ whose data has been corrupted to an ES with correct file.

\end{itemize}

\begin{figure}[!ht]
	\centering
	\includegraphics[width=3.5in]{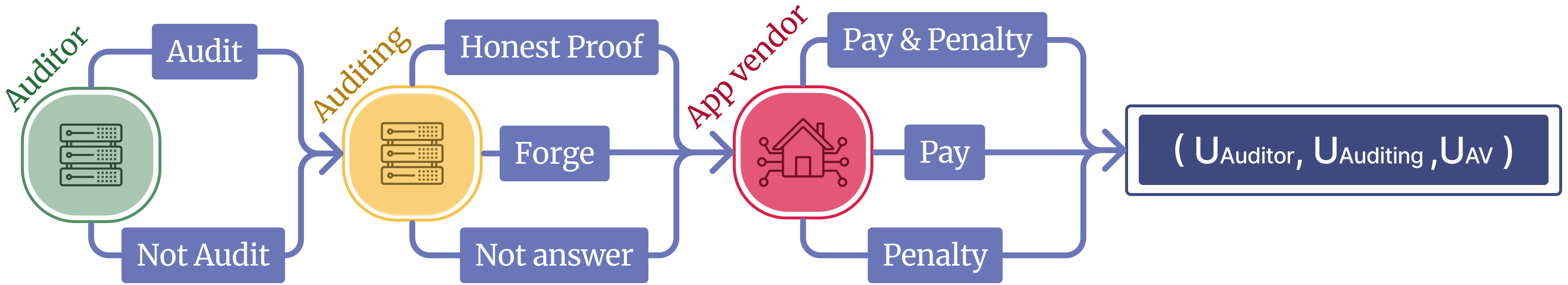}
	\caption{Game tree.} \label{gameTree}
\end{figure}

\section{Game Overview}\label{gta}
\begin{figure}[!ht]
	\centering
	\includegraphics[width=3.5in]{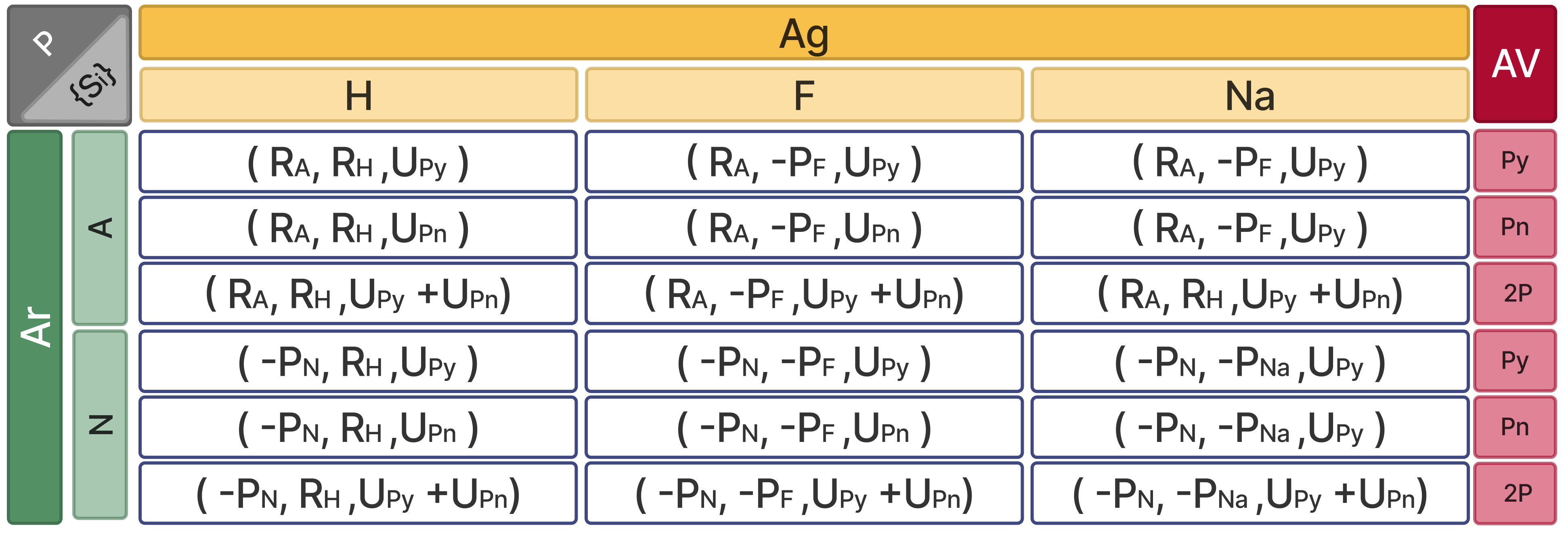}
	\caption{Game table.}
	\label{gameTable}
\end{figure}

In Section \ref{pf}, we described a data auditing system model in which the TPA was removed, and ESs were replaced. But is this possible in reality? To what extent can this TPA be replaced by a distributed data integrity auditing protocol run by the ESs themselves? What makes ESs more motivated than the TPA to audit each other? We will utilize a game theoretic approach to answer these questions in the following. To put it informally, in our target system, ESs are competing with each other to attract more users and satisfy the AV. To this end, compared to TPA, they have more incentive to audit their competitors, find corrupted data, and report it to AV. In fact, the data auditing process is more beneficial for ESs, and they are a more suitable option than TPA for integrity checking. Thereby, in an auditing process, we consider the interaction between the AV, the auditor ES, and the auditing ES in the form of an extensive form game:

\begin{Definition}[FDIA Game] The sequential FDIA game is the triplet $G = ( P , \;  \lbrace S_i \rbrace_{i \in \lbrace P \rbrace }, \; \lbrace U_{i} \rbrace_{i \in \lbrace P \rbrace } )
$, in which the auditor ES is the first, the auditing ES is the second, and the AV is the last player (Figures \ref{gameTree} and \ref{gameTable}). In this game we have\\

The FDIA game is sequential game represented as a triplet $G = ( P , \; \lbrace S_i \rbrace_{i \in \lbrace P \rbrace }, \; \lbrace U_{i} \rbrace_{i \in \lbrace P \rbrace } )$, where the auditor ES goes first, followed by the auditing ES, and finally the AV (Figures \ref{gameTree} and \ref{gameTable}). The game includes:\\

{\small  
$P = \lbrace Ar : \text{Auditor} \in ESs, \, \lbrace Ag: \text{Auditing}  \in ESs, \, AV \rbrace, $

$S_{AV} = \lbrace Py: \text{ Pay}, \, Pn : \text{ Penalize }, \, 2P : \text{ Pay} \& \text{Penalize}\rbrace$ 
	
$S_{Ar} = \lbrace A : \text{ Audit } , N :\text{ No audit } \rbrace$ 
	
$S_{Ag} = \lbrace H: \text{ ‌Be honest } , F: \text{ Forge } , Na: \text{ No answer }\rbrace$ 
\noindent
\begin{table}[ht!]
	\begin{tabular}{ccc}
	\makecell{
	$
	U_{Ar} =
	\begin{cases}
		R_{A}  & A\\
		-P_{N}  & N\\
	\end{cases}
	$}
		&
		\makecell{
		$
U_{Ag} =
\begin{cases}
	R_{H}  & H\\
	-P_{F}  & F\\
	-P_{Na}  & Na\\
\end{cases}
$}
&
		\makecell{
$
	U_{AV} =
	\begin{cases}
		U_{Py}  & Py\\
		U_{Pn}  & Pn\\
		U_{Py} + U_{Pn}  & 2P\\
	\end{cases}
	$
	}
	\end{tabular}
\end{table}
}
\end{Definition}

We now define the Nash equilibrium strategy profile in our FDIA game as follows:

\begin{Definition}\label{NE}
	A Nash equilibrium of the FDIA game is a strategy profile $s^* = (s_{Ar}^{*}, s_{Ag}^{*}, s_{AV}^{*})$ where no player can improve their utility by changing their strategy alone, i.e.,
	\begin{align*} 
		U_n (s_{n}^{*}, s_{-n}^{*}) \leq U_n (s_{n}, s_{-n}^{*}), \, \; \forall s_{n} \in S_{n}, \; n \in  P.
	\end{align*}
\end{Definition}

The Nash equilibrium is a point where users reach a mutually satisfactory solution, and auditing ESs have no incentive to forge. It indicates a strategy profile where all players have no incentive to deviate from their current strategies. NE helps determine if auditor ESs want to participate in the system.

\subsection{Game-theoretic analysis}
In this section, we investigate whether a Nash equilibrium exists for the FDIA game. To do so, initially, we present the notion of best response: 

\begin{Definition}\label{BR}
	Player $n$'s strategy $s^{*}_n \in S_{n}$ is a best response given other players' strategies $s_{-n}$ if
	\begin{equation*} 
		U_n (s_{n}^{*}, s_{-n}) \leq U_n (s_{n}, s_{-n}), \, \; \forall s_{n} \in S_{n}.
	\end{equation*}
\end{Definition}

\begin{figure}[ht!]
	\centering
	\includegraphics[scale=.08]{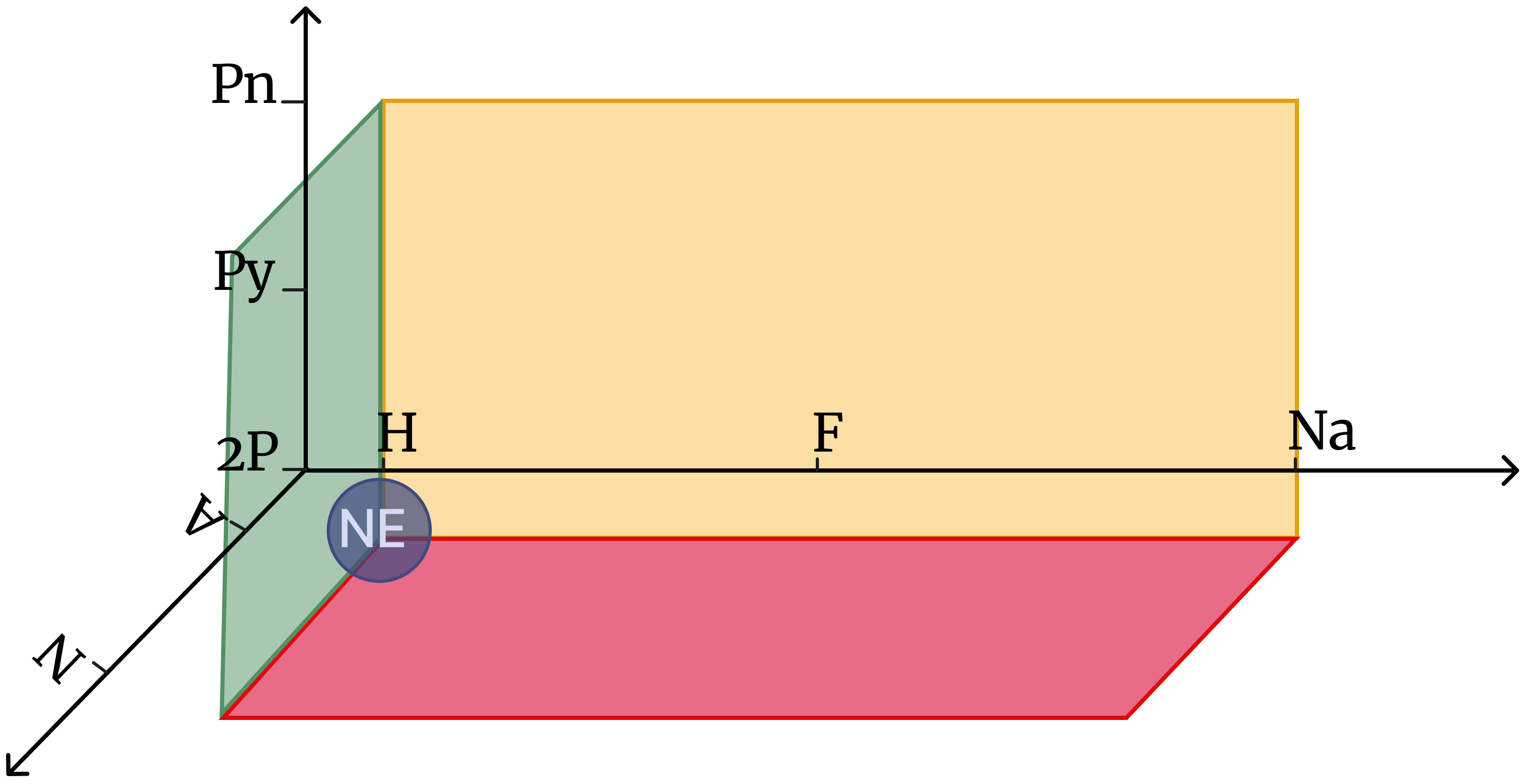}
	\caption{Nash equilibrium daiagram.}
	\label{NEdiagram}
\end{figure}

The Nash equilibrium is achieved when all players adopt their best response, as defined in Definitions \ref{NE} and \ref{BR}. Referring to the table of players' utilities in Fig. \ref{gameTable}, we observe that the auditor ES will choose strategy $A$, the auditing ES will choose strategy $H$, and the AV will choose strategy $2P$ as their best response. Therefore, as depicted in Fig. \ref{NEdiagram}, the strategy profile $( A, \, \, H, \, \, 2P)$ is the Nash equilibrium of the FDIA game. 

\section{Security analysis}\label{sa}
Assume that $\mathcal{A}$ , $\mathcal{C}$, and $\Pi$ are a PPT adversary, a PPT challenger, and our proposed FDIA, respectively. Consider the following experiment. \\
\noindent
\textbf{The proof forging experiment} $Proof-forge_{\mathcal{A},\Pi}(n)$:
\begin{enumerate}
	\item \textbf{Start}: $\mathcal{C}$ considers an initially empty list $L_{id}$, runs $(params, msk) \leftarrow \textbf{Setup}(1^n)$ and returns the public parameters $params$ to $\mathcal{A}$.
	
	\item \textbf{Phase 1}: $\mathcal{A}$ asks a multitude of questions to the oracle $\mathcal{O}_{\textbf{AV.keyGen}}(id)$, and $\mathcal{C}$ answers them as follows:
	\begin{itemize}
		\item $\mathcal{O}_{\textbf{AV.keyGen}}(id)$: Given an identifier $id$ , $\mathcal{C}$ adds $id$ to $L_{id}$, executes $(sk_{id} , s) \leftarrow \textbf{AV.KeyGen}( params, msk, id)$ and returns $(sk_{id} , s)$ to $\mathcal{A}$.
	\end{itemize}
	
	\item \textbf{Phase 2}: When \textbf{Phase 1} is finished, $\mathcal{A}$ sends a challenge identifier $id^{*}$ to $\mathcal{C}$. If $id^{*} \in L_{id}$, $\mathcal{C}$ asks $\mathcal{A}$ to choose another challenge identifier. Otherwise, $\mathcal{C}$ executes $(sk^{*}_{id}, s^{*}) \leftarrow \textbf{AV.keyGen}(params, msk, id^{*})$. Then, $\mathcal{A}$ makes a polynomial number of queries to the oracles $\mathcal{O}_{\textbf{AV.keyGen}}(id^{*})$ and $\mathcal{O}_{\textbf{TagGen}}(F)$. The challenger $\mathcal{C}$ responds to them as follows:
	\begin{itemize}
		\item $\mathcal{O}_{\textbf{AV.keyGen}}(id)$: For an identifier $id$ from $\mathcal{A}$, if $id \neq id^{*}$, $\mathcal{C}$ runs $(sk_{id}, s) \leftarrow \textbf{AV.keyGen}(params, msk, id)$ and outputs $(sk_{id}, s)$ to $\mathcal{A}$. 
		
		\item $\mathcal{O}_{\textbf{TagGen}}(F)$: For a file $F$, $\mathcal{C}$ executes $F_{t} \leftarrow \textbf{TagGen}(params, F, (sk^{*}_{id}, s^{*}))$ and gives $F_{t}$ to $\mathcal{A}$.
	\end{itemize} 
	
	\item \textbf{Challenge}: $\mathcal{C}$ selects $\hat{F} = \hat{f}_1 || ... ||\hat{f}_m \in \mathbb{Z}_{p}^{m}$, $i \in \lbrace 1, ..., m \rbrace $, and $x \in \mathbb{Z}_p$ randomly and considers 
	\begin{align*}
\hat{F}' = \hat{f}_1 || . . .  || \hat{f}_{i-1} ||x|| \hat{f}_{i+1} || . . . || \hat{f}_m.
	\end{align*} 
	
	Then $\mathcal{C}$ executes $\hat{F}_{t} \leftarrow \textbf{TagGen} ( params, \hat{F}, (sk^{*}_{id}, s^{*}))$. Also, it selects an identifier $\hat{id}$ and executes $sk_{\hat{id}} \leftarrow \textbf{ES.KeyGen}(params, msk, \hat{id})$. Afterward, it runs $ch_{\hat{F}} = (\hat{\alpha}, \hat{\beta}, I, k_{PRF}, \sigma_{k_{PRF}}, \hat{id}) \,  \leftarrow \, \textbf{Challenge}( params, m, \hat{id}, sk_{\hat{id}})$. Finally, $\mathcal{C}$ returns $(\hat{t} = \lbrace \hat{t}_{i} \rbrace_{i = 1}^{m}, name_{\hat{F}}, \hat{F}', \hat{\alpha}, \hat{\beta})$ to $\mathcal{A}$.
	
	\item \textbf{Forge}: $\mathcal{A}$ returns a proof $\hat{m'}$  to $\mathcal{C}$. 
\end{enumerate}

\begin{Definition}
	We say that our FDIA scheme $\Pi$ is secure if for any PPT adversary $\mathcal{A}$ the value of $Pr[Proof-forge_{\mathcal{A},\Pi}(n) = 1]$ is a negligible function in $n$.
\end{Definition}

\begin{Theorem}
	If the hardness assumption of the BDH problem with respect to $\mathcal{G}$ is held, then our FDIA scheme remains secure in the random oracle model.
\end{Theorem}

\begin{proof} 
	Assume $\mathcal{A}$ is a PPT adversary, $n$ a security parameter, and $\Pi$ our FDIA scheme. We show that the value of $Pr[Proof-forge_{\mathcal{A},\Pi}(n) = 1]$ is a negligible function in $n$, where $Proof-forge_{\mathcal{A},\Pi}(n)$ is the described proof forging experiment. Consider a PPT distinguisher $\mathcal{D}$ that takes a tuple $(n, p, G, G' , e, g, g^a, g^b, g^c )$ and aims to calculate $e(g, g)^{abc}$, where $(p, G, G', e) \leftarrow \mathcal{G}(1^n)$ , $g \in G$ is a random generator, and $a,b,c \in \mathbb{Z}_{p}$ are chosen uniformly at random. To solve the BDH problem, $\mathcal{D}$ runs $\mathcal{A}$ as a subroutine as follows:
	
	\begin{enumerate}
		\item \textbf{Start}: 
		$\mathcal{D}$ pickes $H_1, H_2 : \lbrace 0, 1\rbrace ^{*} \rightarrow G$, $PRF : \mathbb{Z}_p \times \mathbb{Z}_p \rightarrow \mathbb{Z}_p$, IDAS scheme $\Pi=(\textbf{Setup}_{sign}, \textbf{KeyGen}_{sign}, \textbf{Sign},\textbf{Agg}, \\ \textbf{Vrfy})$, and $m \, \in \, \mathbb{N}$. It executes $(pk_{sign} , msk_{sign}) \leftarrow \textbf{Setup}_{sign}(1^n)$ and gives $params = (n, p,  g, G, G', e, h = g^a, H_1, H_2, PRF, m, \Pi, pk_{sign})$ to $\mathcal{A}$.
		\item \textbf{Phase 1}: $\mathcal{A}$ asks scores of queries to the oracles $\mathcal{O}_{H_1}$, $\mathcal{O}_{H_2}$, and  $\mathcal{O}_{\textbf{AV.keyGen}}(id)$, and $\mathcal{D}$, considering two initially empty sets $E_{H_1}$, $E_{H_2}$,answers as follows:
		\begin{itemize}
			\item $\mathcal{O}_{H_1}(X)$ : Given $X \in \lbrace 0 , 1 \rbrace^*$, $\mathcal{D}$ checks whether $E_{H_1}$ includes $(X, x)$ or not. If not, it selects $a \in \mathbb{Z}_p$ uniformly at random and adds $(X, x)$ to $E_{H_1}$. Finally, it returns $H_1(X) = (g^a)^x$.
			\item $\mathcal{O}_{H_2}(X)$ : Again, given $X \in \lbrace 0 , 1 \rbrace^*$, if $E_{H_2}$ does not include $(X, x)$, $\mathcal{D}$ selects $x \in \mathbb{Z}_p$ randomly, adds $(X, x)$ to $E_{H_2}$, and returns $H_2(X) = (g^a)^x$.
			\item $\mathcal{O}_{\textbf{AV.keyGen}}(id)$: For an identifier $id$ ,if $E_{H_1}$ does not include $(id, x_{id})$, $\mathcal{D}$ selects $x_{id} \in \mathbb{Z}_p$ uniformly at randomly and adds $(id, x_{id})$ to $E_{H_1}$. Afterward it computes the values $s = H_{1}(id)^{\eta} = (g^a)^{x_{id}\eta}$, $sk_{id} \leftarrow \textbf{KeyGen}_{sign}(pk_{sign}, msk_{sign}, id)$. Finally, $\mathcal{D}$ gives $(sk_{id}, s)$ to $\mathcal{A}$.
		\end{itemize}
		\item \textbf{Phase 2}: $\mathcal{A}$ picks an identifier $id^{*}$ such that the secret key associated with $id^*$ has not been queried in \textbf{Phase 1}. Then $\mathcal{D}$ computes $sk^{*}_{id} \leftarrow \textbf{KeyGen}_{sign}(pk_{sign}, msk_{sign}, id^{*})$ and $s^{*} = H_{1}(id)^z$. Also it selects a random $r \in \mathbb{Z}_p$ and calculates $h' = g^r$ and $h" = h^r$. 
		Afterward, $\mathcal{A}$ poses a polynomial number of queries to the oracles listed below, and $\mathcal{D}$ responds to $\mathcal{A}$ as follows. Note that according to the notations used in Fig.\ref{algorithms}, we have $r = b$, and thus $h" = h^b = g^{ab}$.
		\begin{itemize}
			\item $\mathcal{O}_{H_1}(X)$ and $\mathcal{O}_{H_2}(X)$ : $\mathcal{D}$ answers these oracles the same as in \textbf{Phase 1}.
			\item $\mathcal{O}_{\textbf{AV.keyGen}}(id)$: If $id \neq id^{*}$, then $\mathcal{D}$ generates the requested keys the same as in \textbf{Phase 1}.
			\item $\mathcal{O}_{\textbf{TagGen}}(F)$: For a file $F = (f_1 , ... , f_m)$, $\mathcal{D}$ first selects $name_F \in \lbrace 0, 1 \rbrace^{*}$ and uniform elements $t_1, ..., t_m \in G$. Then, it returns $(name_F , t=\{t_i \}_{i=1}^{m})$. 
			It can be observed that assuming $H_2(name_F g^{ab} i) = (t_i . s^{-f_i})^{-b}$ for $i = 1, ..., m$, then $t$ is a valid tag. 
		\end{itemize} 
		\begin{Remark} \label{remark11} 		
		It is important to highlight that according to the hardness assumption of the BDH problem, and subsequently the DH problem, the adversary $\mathcal{A}$, except with a negligible probability, does not query an element in the form of $X = (name_F ||g^{ab} ||i)$ to the oracle $\mathcal{O}_{H_2}$. Therefore, based on the programmability feature of the random oracle model \cite{sec}, we can presume that $H_2 (name_F ||g^{ab} ||i) = (t_i . s^{-f_i})^{-b}$.
		\end{Remark}
		\item \textbf{Challenge}: First, $\mathcal{D}$ chooses $\hat{i} \in	\{1, . . . , m\}$ , $x \in \mathbb{Z}_p$ , and $I \subset \{1, . . . , m\}$ uniformly at random. Then, it considers an unknown data file $\hat{F} = \hat{f}_1 || ...||\hat{f}_{i-1} ||b||\hat{f}_{i+1} || ... ||\hat{f}_m$, where $\hat{f}_j \in \mathbb{Z}_p$ is selected uniformly at random for $j \in \{1, . . . , m\} \setminus \{\hat{i}\}$. It also sets $\hat{F}' = \hat{f}_1 || . . . ||\hat{f}_{i-1} ||x||\hat{f}_{i+1} || . . . ||\hat{f}_m$, $\Lambda = e(H_1(id), h)$, $\alpha = g^{\lambda}$, and $\beta = \Lambda ^{\lambda}$ for a uniformly at random selected $\lambda \in \mathbb{Z}_{p}$ and returns $(\hat{t} = \{\hat{t}_{i}\}_{i=1}^{m}, name_{\hat{F}}, \hat{F}', \alpha , \beta , k_{PRF})$ to $\mathcal{A}$, where $\hat{t}$ is the tag associated with $\hat{F}$. Note that according to the notations employed in Fig. \ref{algorithms}, we have $\lambda = c$ .
		\item \textbf{Forge}: $\mathcal{A}$ returns a proof $ \hat{m'}$ to $\mathcal{D}$.
	\end{enumerate}
	Upon receiving $\hat{m'}$, the challenger $\mathcal{C}$ first computes 
	
	{\footnotesize
	\begin{align*}
	w = e(t_{\hat{i}}^{f_{k_{PRF}(\hat{i})}} , g^{bc}) . \prod\limits_{i \in I \setminus \{\hat{i} \}} e(H_{2}(name_F || h" || i)^{f_{k_{PRF}(i)}} , h'^{c})
	\end{align*}
	}
	and outputs $(\hat{m'} w^{-1})^{-(x_{id}\eta f_{k_{PRF}(\hat{i})})}$. Note that as we mentioned in Remark \ref{remark11}, we have assumed that $H_2 (name_F ||g^{ab} ||\hat{i}) = (t_{\hat{i}} . s^{-b})^{-r}$. Therefore, if $\mathcal{A}$ succeeds in forging an integrity proof, then we have 
		
	{\scriptsize	
	\begin{align*}
		& \hat{m'} = \prod_{i \in I} e(H_2(name_F||h"||i)^{f_{k_{PRF}}(i)} , h'^{\lambda}) \\
		& = \prod_{i \in I \setminus \{i^{*}\}} e(H_2(name_F||h"||i)^{f_{k_{PRF}}(i)} , h'^{\lambda}) .  e(H_2(name_F||h"||\hat{i})^{f_{k_{PRF}}(\hat{i})} , h'^{\lambda}) \\
		& = \prod_{i \in I \setminus \{\hat{i}\}} e(H_2(name_F||h"||i)^{f_{k_{PRF}}(i)} , h'^{\lambda}) .  e((t_{\hat{i}} . s^{-b})^{-r f_{k_{PRF}}(\hat{i})} , g^{bc})\\
		& = \prod_{i \in I \setminus \{\hat{i}\}} e(H_2(name_F||h"||i)^{f_{k_{PRF}}(i)} , h'^{\lambda}) .  e((t_{\hat{i}} . H_{1}(id)^{\eta b})^{-b r f_{k_{PRF}}(\hat{i})} , g^{bc})\\
		& = \prod_{i \in I \setminus \{\hat{i}\}} e(H_2(name_F||h"||i)^{f_{k_{PRF}}(i)} , h'^{\lambda}) .  e((t_{\hat{i}} .(g^{a})^{x_{id}\eta b})^{-b r f_{k_{PRF}}(\hat{i})} , g^{bc})\\
		& = \prod_{i \in I \setminus \{\hat{i}\}} e(H_2(name_F||h"||i)^{f_{k_{PRF}}(i)} , h'^{\lambda}) . e(t_{\hat{i}}^{-b r f_{k_{PRF}}(\hat{i})} , g^{bc}). e( (g^{a})^{x_{id}\eta r f_{k_{PRF}}(\hat{i})} , g^{bc})\\
		& = \prod_{i \in I \setminus \{\hat{i}\}} e(H_2(name_F||h"||i)^{f_{k_{PRF}}(i)} , h'^{\lambda}) . e(t_{\hat{i}}^{-b r f_{k_{PRF}}(\hat{i})} , g^{bc}) . e(g^{a}, g^{bc}) ^{x_{id}\eta r f_{k_{PRF}}(\hat{i})}	\\
		& = w.(e(g,g)^{abc})^{x_{id}\eta r f_{k_{PRF}}(\hat{i})}
	\end{align*}}
	Therefore, $(m'w^{-1})^{-(x_{id}\eta r f_{k_{PRF}}(\hat{i}))} = e(g, g)^{abc}$, and thus
	{\footnotesize	
	\begin{align*}
	Pr[Proof-forge_{\mathcal{A},\Pi}(n) = 1] \leqslant Pr[\mathcal{D}(n, p, G, G' , e, g, g^a, g^b, g^c ) = e(g, g)^{abc}].
	\end{align*}
	}
		
	On the other hand, by the hardness assumption of the BDH problem, we have $Pr[(n, p, G, G' , e, g, g^a, g^b, g^c ) = e(g, g)^{abc} ]$ is a negligible function in $n$. Thus, $Pr[Proof-forge_{\mathcal{A},\Pi}(n) = 1]$ is also a negligible function. This completes the proof.
\end{proof}
\section{Performance analysis}\label{per}
In this section, to evaluate the efficiency of our FDIA scheme, we compare its performance in terms of execution time and storage overhead with the approaches outlined in \cite{int8}, \cite{rw2}, \cite{rw3}, \cite{rw4}, and \cite{rw7}. We present the comparison results using both practical and theoretical analysis. In practical testing, we implemented our proposed scheme to create a prototype system. Using this prototype system, we primarily assess and analyze the execution time and data transmission size of the aforementioned schemes under identical conditions.

\subsection{Theoretical analysis}
Tables \ref{performance} and \ref{performance2} display our asymptotic analysis, where we accurately determine the computation and storage overheads incurred when using FDIA and the methods presented in mentioned studies. To compute the execution time overhead ($\mathcal{O}_{1}$ to $\mathcal{O}_{4}$), we considered the operations described in the caption of Table \ref{performance}. Then, we calculated storage costs ($\mathcal{O}_{5}$ to $\mathcal{O}_{7}$) based on the size of elements mentioned in the caption of Table \ref{performance2}. In Tables \ref{performance} and \ref{performance2}, $N$, $m$, and $| I |$ denote the number of data files, blocks, and challenged blocks, respectively.

\begin{table*}[!ht]
			\caption{
				\begin{footnotesize}
					Comparison of various schemes computational overhead.
				\end{footnotesize}
			}
				\scalebox{0.8}{
				\begin{tabular}{|c|c|c|c|c|c|c|}
					\hline
					& Ours & B. Li \cite{int8} & T. Shang \cite{rw2} & J. Shu \cite{rw3}& Y. Li \cite{rw4}& M. Tian \cite{rw7} \\ 
					\hline
					\hline
					$\mathcal{O}_{1}$ 
					& \makecell{$m(2T_{e_{1}} + T_{H}) +  2T_{e_{1}}$ \\ $  + T_{sign} + T_{rand_{Z}}$}
					& \makecell{--}
					& \makecell{$m(2T_{e_{1}} + T_{H}) + 2T_{e_{1}}  + T_{sign} $ \\ $+ T_{rand_{Z}} + T_{H}$}
					& \makecell{$m(3T_{e_{1}} + 2T_{H}) + T_{e_{1}}$}
					& \makecell{$m(2T_{e_{1}} + T_{H}) + 3T_{e_{1}} + 2T_{rand_{Z}} + T_{H}$}
					& \makecell{$m(T_{A} + T_{H} + T_{M_Z}) $ \\ $ + 2T_{e_{1}} $}\\
					\hline
					$\mathcal{O}_{2}$ 
					& \makecell{$N(T_{rand_{Z}} + T_{sign}) +  $ \\ $ 2T_{e_1}$}
					& \makecell{$2NT_{rand_{Z}}$}
					& \makecell{$N(|I|T_{rand_{Z}} + T_{H} + T_{p} $ \\ $ + T_{e_1} + T_{e_2})$}
					& \makecell{$N|I|T_{rand_{Z}}$}
					& \makecell{$N|I|T_{rand_{Z}}$}
					& \makecell{$N|I|T_{rand_{Z}}$}\\
					\hline
					$\mathcal{O}_{3}$ 
					& \makecell{$|I|(T_{e_1} + T_{M_Z}) + $ \\ $ T_{p} + T_{e_2}$}
					& \makecell{$N(T_{rand_Z} + T_{H}  + T_{e_1} $ \\$+ 2 T_{G_1})$}
					& \makecell{$|I|(T_{e_1}+T_{M_Z}) + T_{p}  + T_{e_1} $\\$ + 2T_{H}$}
					& \makecell{$|I|(2T_{e_1} + T_{e_1} + T_{H})$}
					& \makecell{$|I|(T_{e_1}+T_{M_Z})$}
					& \makecell{$|I|(3T_{H} + T_{e_1})$}\\
					\hline
					$\mathcal{O}_{4}$ 
					& \makecell{$|I|(T_{e_1} + T_{p} + T_{H})$}
					& \makecell{$ T_{H} +  3T_{e_1} + 2 T_{M_Z}$}
					& \makecell{$|I|(T_{e_1} + T_{p} + T_{H}) + T_{e_1}$}
					& \makecell{$2|I|(T_{e_1} +T_{H}) + 3(T_{p} + T_{e_1})$}
					& \makecell{$|I|(2(T_{e_1} +T_{H}) + T_{M_Z} +  T_{G_1}) + 3T_{p}  $ \\$ +2T_{H} + T_{e_1} $}
					& \makecell{$|I|(3T_{H} + T_{e_1})$}\\
					\hline
					\hline
				\end{tabular}
				}
				\\
				
				\scalebox{0.8}{
				Notes. $\mathcal{O}_{1}$: Tag generation time,
				$\mathcal{O}_2$: Challenge generation time, 
				$\mathcal{O}_3$: Proof generation time, 
				$\mathcal{O}_4$: Proof Check time, 
				$T_p$: Pairing operation time,
				$T_{e}$ , $T_{e'}$: Exponential operation in
				}
				\scalebox{0.8}{
				$G$ and $G'$,
				$T_{sign}$: Signature generation time,
				$T_{vrfy}$: Signature verification time,
				$T_H$: Hashing operation time,
				$T_{M_z}$: Multiplication operation in $\mathbb{Z}_p$,
				$T_{rand_z}$: Random selection of an
				}
				\scalebox{0.8}{
				element from $\mathbb{Z}_p$,
				}
			\label{performance}
\end{table*}

\begin{table}[!ht]
\centering
	\caption{
		\begin{footnotesize}
					Comparison of various schemes storage overhead.
		\end{footnotesize}
	}
	\scalebox{.9}{
		\begin{tabular}{|c|c|c|c|c|c|c|}
			\hline
			& ours & B. Li \cite{int8} & T. Shang \cite{rw2} & J. Shu \cite{rw3}& Y. Li \cite{rw4}& M. Tian\cite{rw7} \\ 
			\hline
			\hline
			 $\mathcal{O}_5$ 
			& \makecell{$ml_{G} + l_{S}$}
			& \makecell{--}
			& \makecell{$ml_{G} + l_{S}+2l_{H}$}
			& \makecell{$2ml_{G}$}
			& \makecell{$(m+3)l_{G} + 3l_{\mathbb{Z}}$}
			& \makecell{$4ml_{G}$}\\
			\hline
			 $\mathcal{O}_6$
			& \makecell{$2l_{G} + l_{S} + l_{\mathbb{Z}}$}
			& \makecell{$l_{G} + l_{\mathbb{Z}}$}
			& \makecell{$l_{G} + l_{G'} + |I|l_{\mathbb{Z}}$}
			& \makecell{$|I|l_{\mathbb{Z}}$}
			& \makecell{$|I|l_{\mathbb{Z}}$}
			& \makecell{$2|I|l_{\mathbb{Z}}$}\\
			\hline
			 $\mathcal{O}_7$ 
			 & \makecell{$l_{G} + l_{S}$}
			 & \makecell{$3l_{G}$}
			 & \makecell{$l_{G} + 2l_{H}$}
			 & \makecell{$3l_{G} + l_{\mathbb{Z}}$}
			 & \makecell{$3l_{G} + l_{\mathbb{Z}}$}
			 & \makecell{$4l_{G}$} \\ 
			\hline
			\hline
			
							\multicolumn{7}{l}{
Notes. $\mathcal{O}_5$: Tag Size,
		$\mathcal{O}_6$: Challenge Size, 
		$\mathcal{O}_7$: Proof Size,
		$l_{\mathbb{Z}}$,  $l_{G}$, $l_{G'}$: Size of elements in 
		$\mathbb{Z}_p$, $G$, and 
		}\\
		\multicolumn{7}{l}{		
		$G'$,
		$l_{H}$: Size of outputs of  $H$,
		$l_{S}$: Size of a signature.
}
		\end{tabular}
		}
		\\

	\label{performance2}
\end{table}

\subsection{Practical analysis}
To evaluate the actual performance of the aforementioned schemes, we conducted experimental simulations on the real-world Amazon Product dataset containing information on various products sold on Amazon \footnote{https://www.kaggle.com/datasets/piyushjain16/amazon-product-data}. The experiments were performed on an Ubuntu 20.04 laptop equipped with an Intel Core i5 Processor, utilizing the hashlib \cite{per2} and python Pairing-Based Cryptography (pyPBC) \cite{per1} libraries. The PBC Library uses Type A denoted as $E(\mathbb{F}_{q}): y^2 = x^3 + x$, where the groups G and G' of order $p$ are subgroups of $E(\mathbb{F}_{q})$, and $p$ and $q$ are equivalent to 160 bits and 512 bits, respectively. We performed experimental tests 100 times on a subset of 10,000 files chosen from the Amazon Product dataset.

Charts a to d of Fig. \ref{keyCharts} show the actual running time of the challenge generation algorithm, providing valuable insights into the efficiency of our scheme compared to schemes \cite{rw2}, \cite{rw3}, \cite{rw4}, and  \cite{rw7}. Remarkably, our scheme demonstrates a significant improvement, performing almost 100 times faster. This is evident from Table \ref{performance}, which indicates that the execution time of our challenge generation algorithm remains unaffected by the value of $|I|$, unlike the other four schemes. Charts e to h of Fig. \ref{keyCharts} illustrate the time of the proof generation algorithm in the mentioned schemes. Here, s represents the number of proven blocks already cached in the auditing's memory. These charts demonstrate that our scheme not only outperforms other schemes in normal mode but also exhibits superior performance compared to itself when there are proven blocks in its memory. 

The comparison of a tag, a challenge, and a proof size is demonstrated in charts a, b, and c of Fig. \ref{keyCharts2}. The chart b presents that our scheme surpasses schemes \cite{rw2}, \cite{rw3}, \cite{rw4}, and  \cite{rw7} in terms of challenge size (almost 100 times faster). This is because, unlike the other four schemes shown in Table \ref{performance2}, our scheme does not depend on the value of $|I|$ for the challenge size. Finally, as from chart d of Fig. \ref{keyCharts2}, our FDIA scheme demonstrates a significantly higher level of efficiency compared to the scheme \cite{int8}. The scheme proposed in \cite{int8}  necessitate the inclusion of complete data files in the verification process, which consequently imposes the burden of local data file storage upon AVs. This, in turn, leads to a substantial increase in storage costs. Fortunately, FDIA effectively addresses this predicament by eliminating the need for local data file storage, thereby resolving the storage cost issue successfully.

\begin{figure*}
	\begin{tabular}{cccc}
		\includegraphics[width=1.66in]{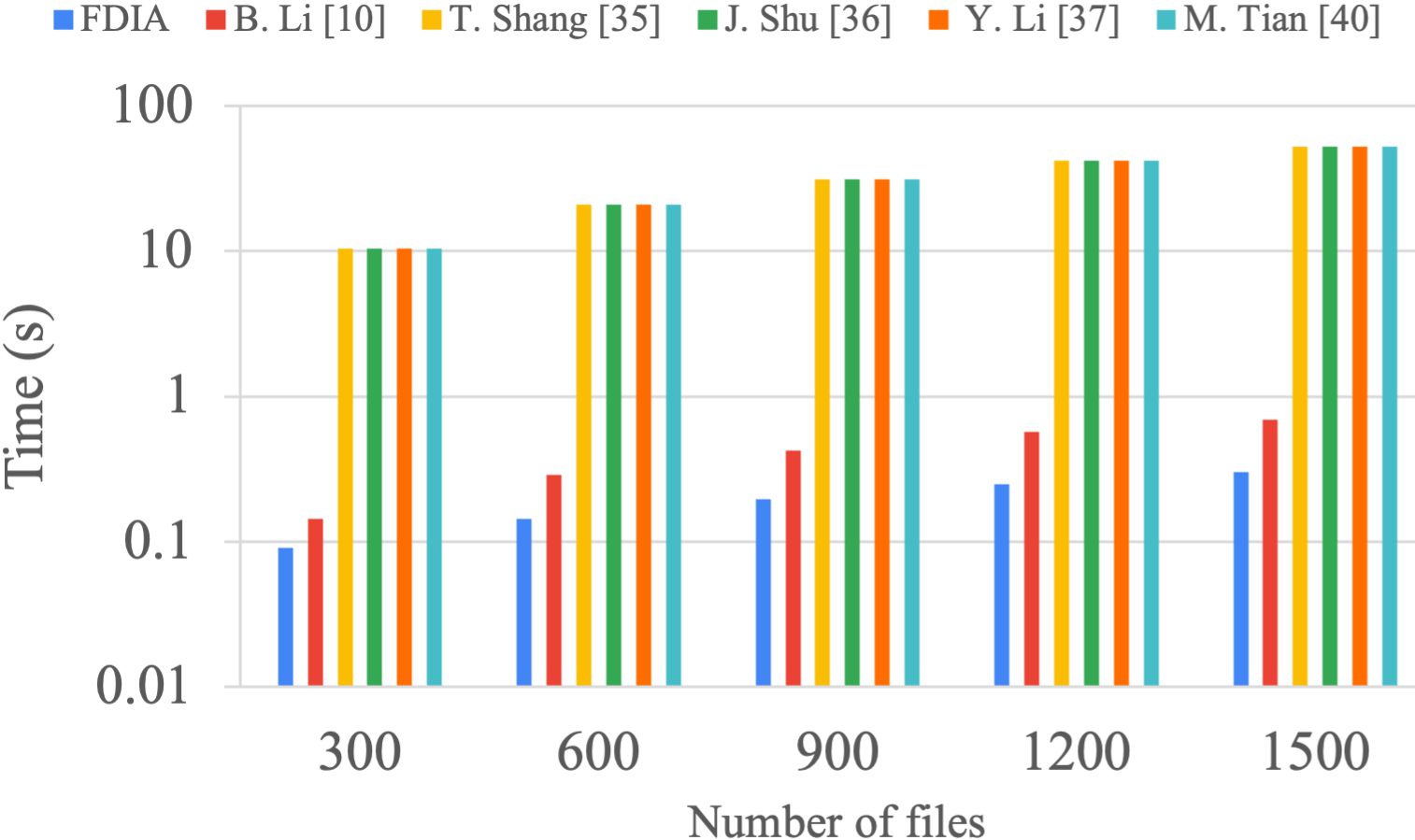}& 
		\includegraphics[width=1.66in]{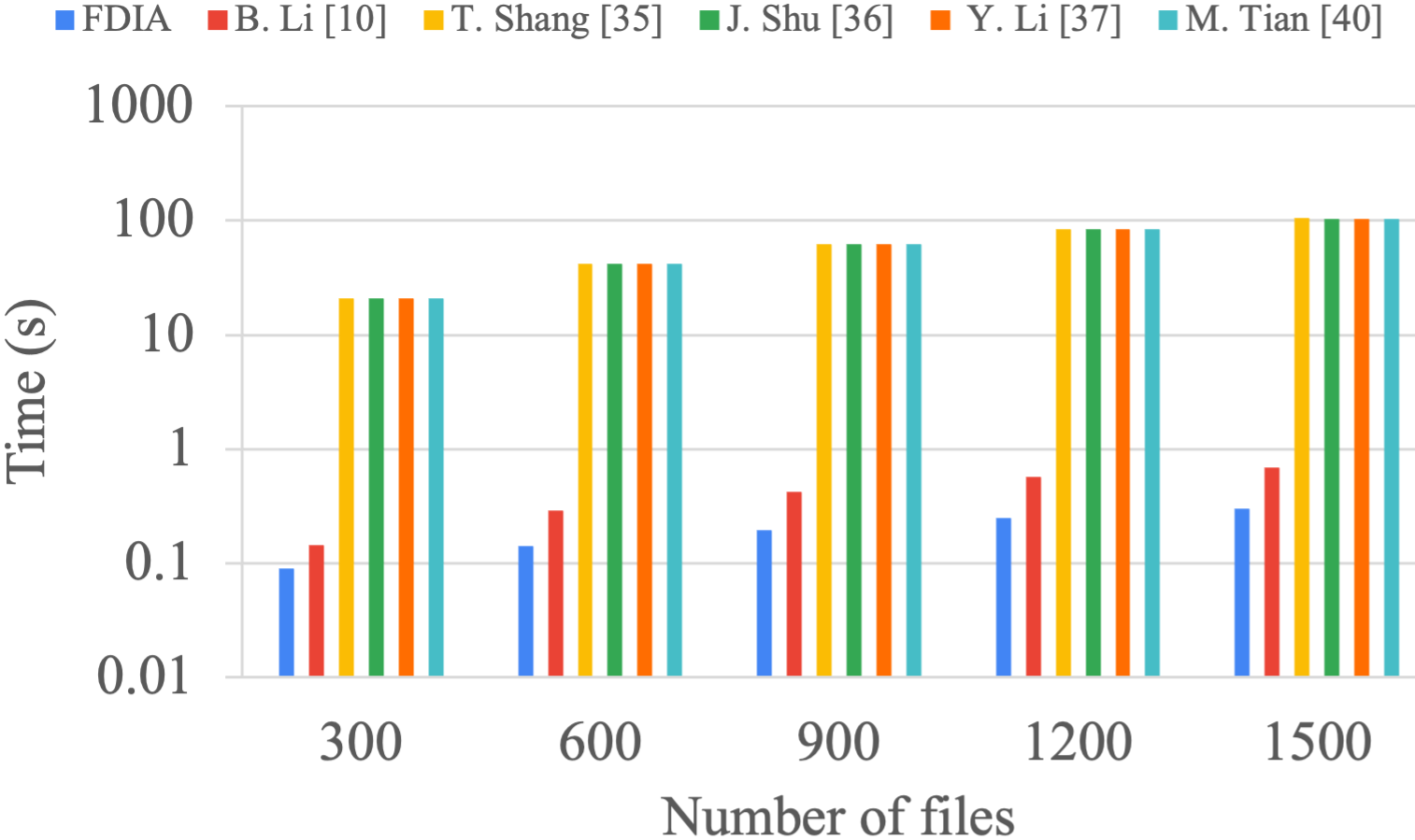}&
		\includegraphics[width=1.66in]{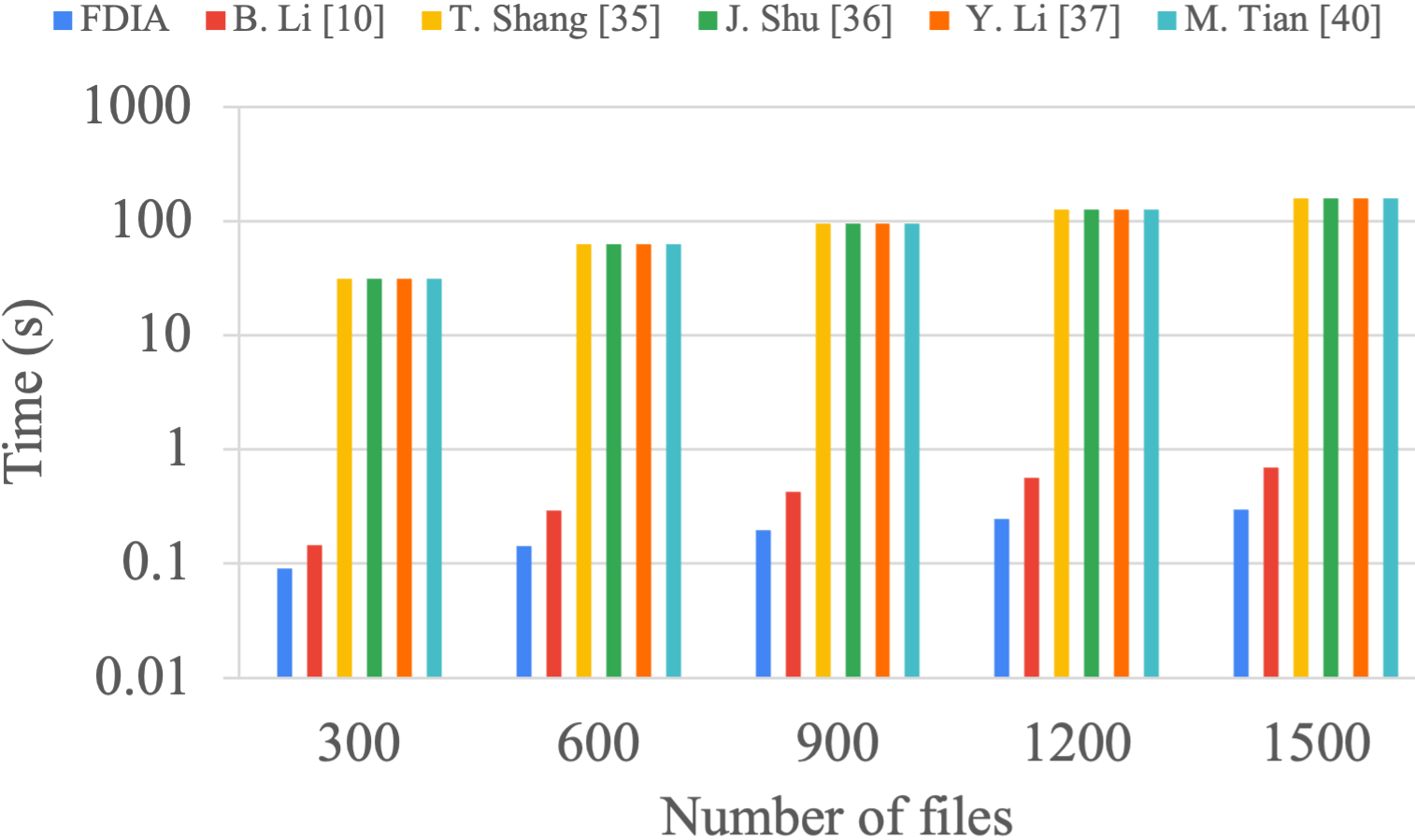}&
		\includegraphics[width=1.66in]{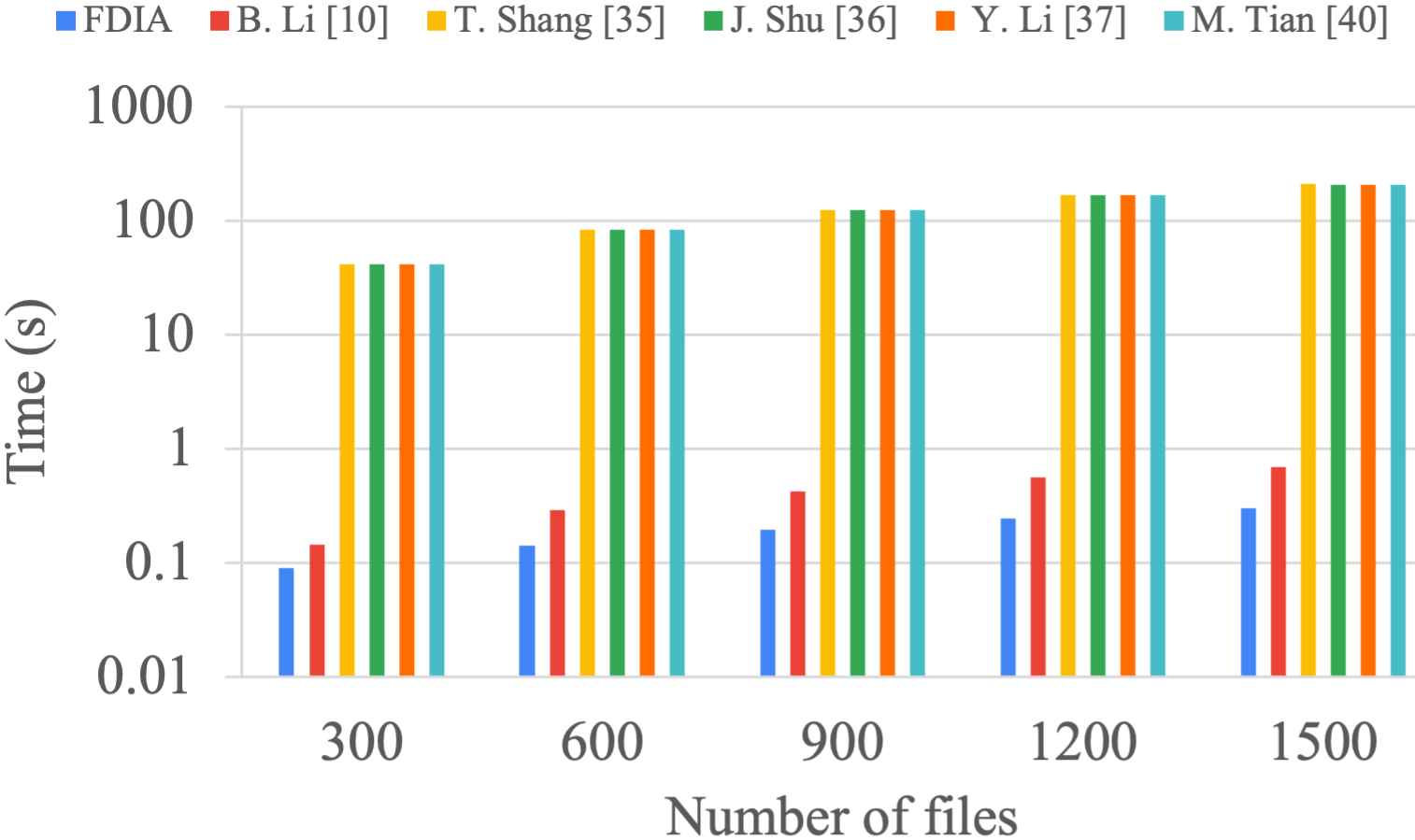}\\
		\scalebox{.5}{(a). Challenge generation time ($| I | =200$).} &
		\scalebox{.5}{(b).  Challenge generation time ($| I | =400$). } &
		\scalebox{.5}{(c).  Challenge generation time ($| I | =600$).} &
		\scalebox{.5}{(d).  Challenge generation time ($| I | =800$). }\\
		\\

		\includegraphics[width=1.65in]{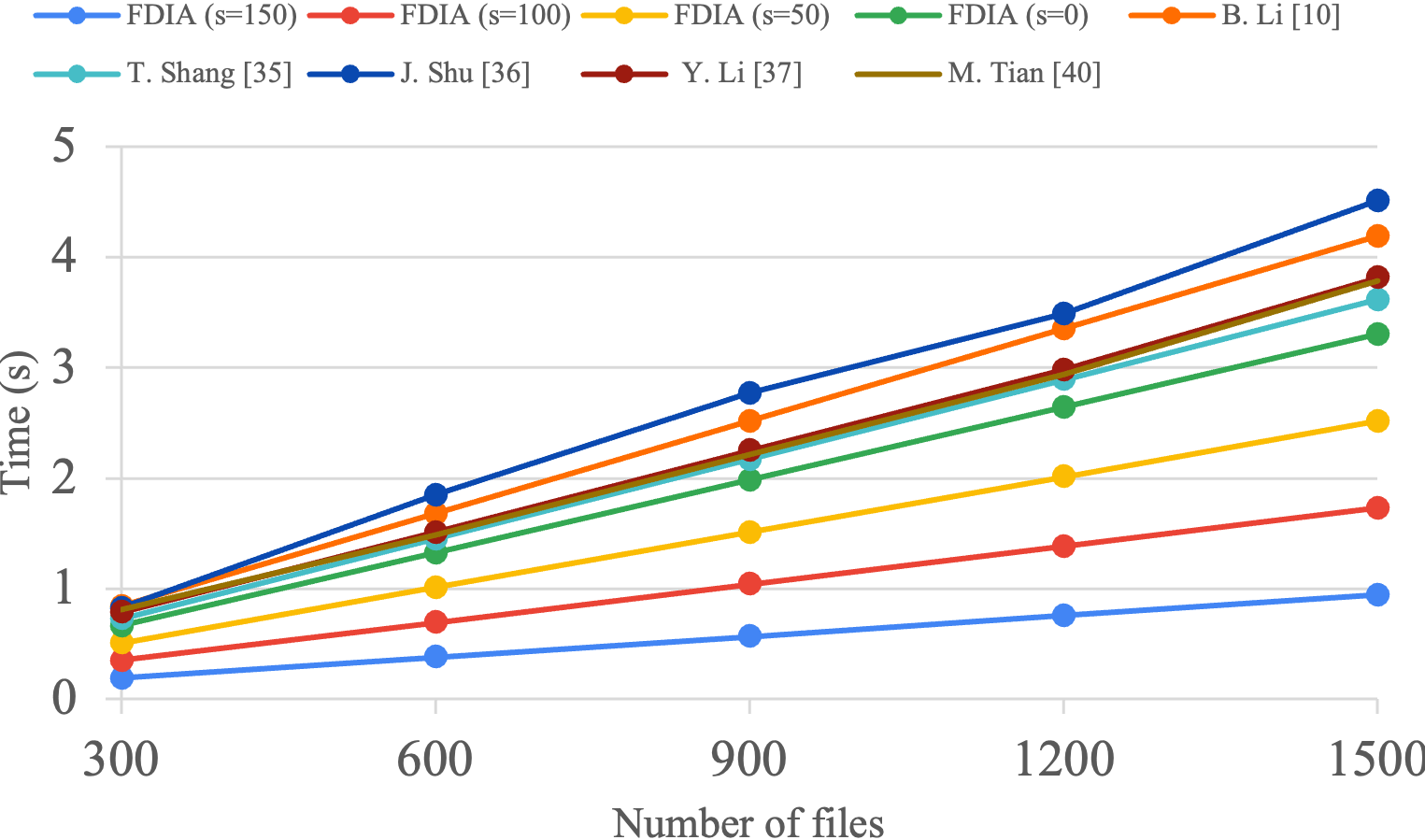}&
		\includegraphics[width=1.65in]{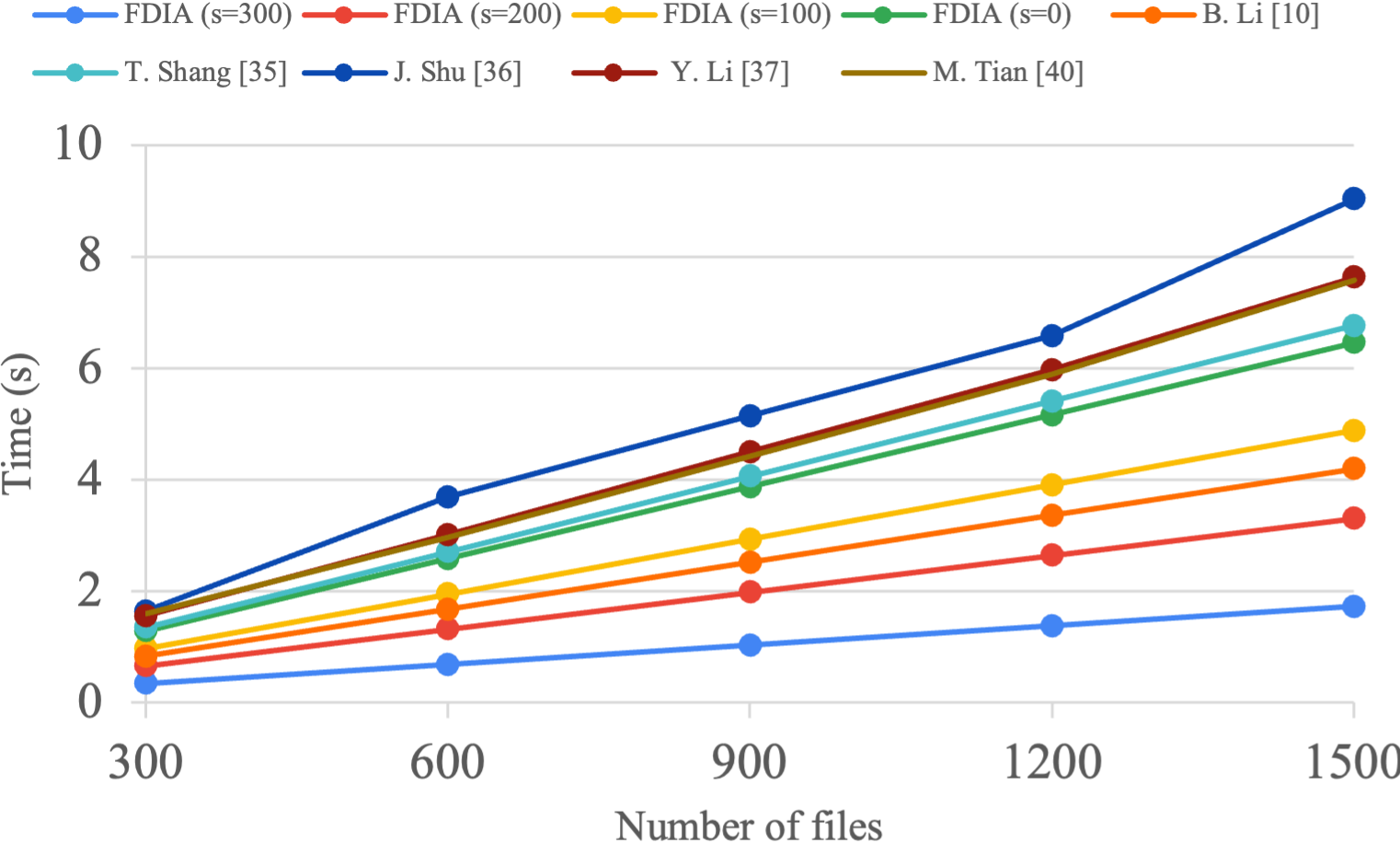}&
		\includegraphics[width=1.65in]{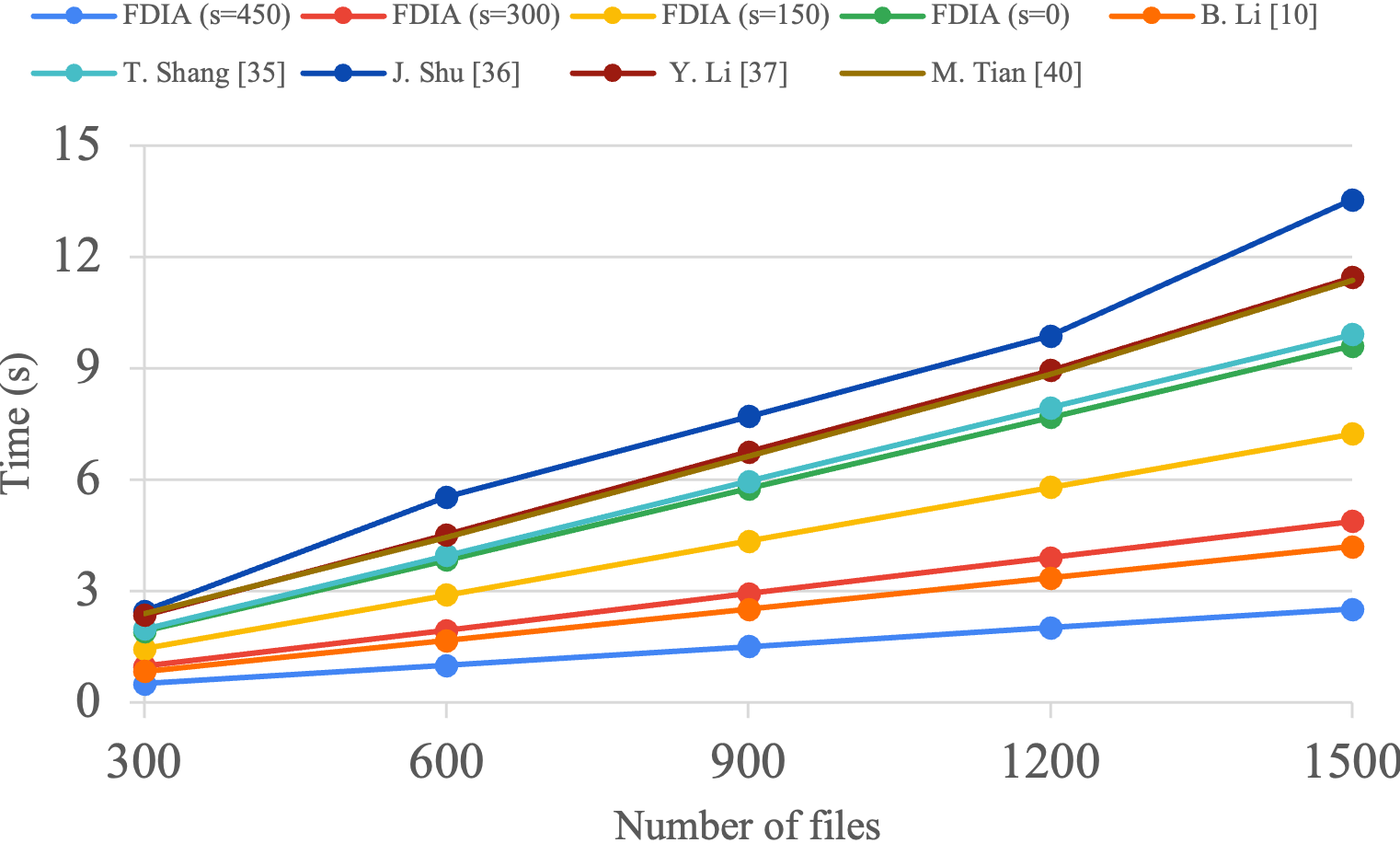}&
		\includegraphics[width=1.65in]{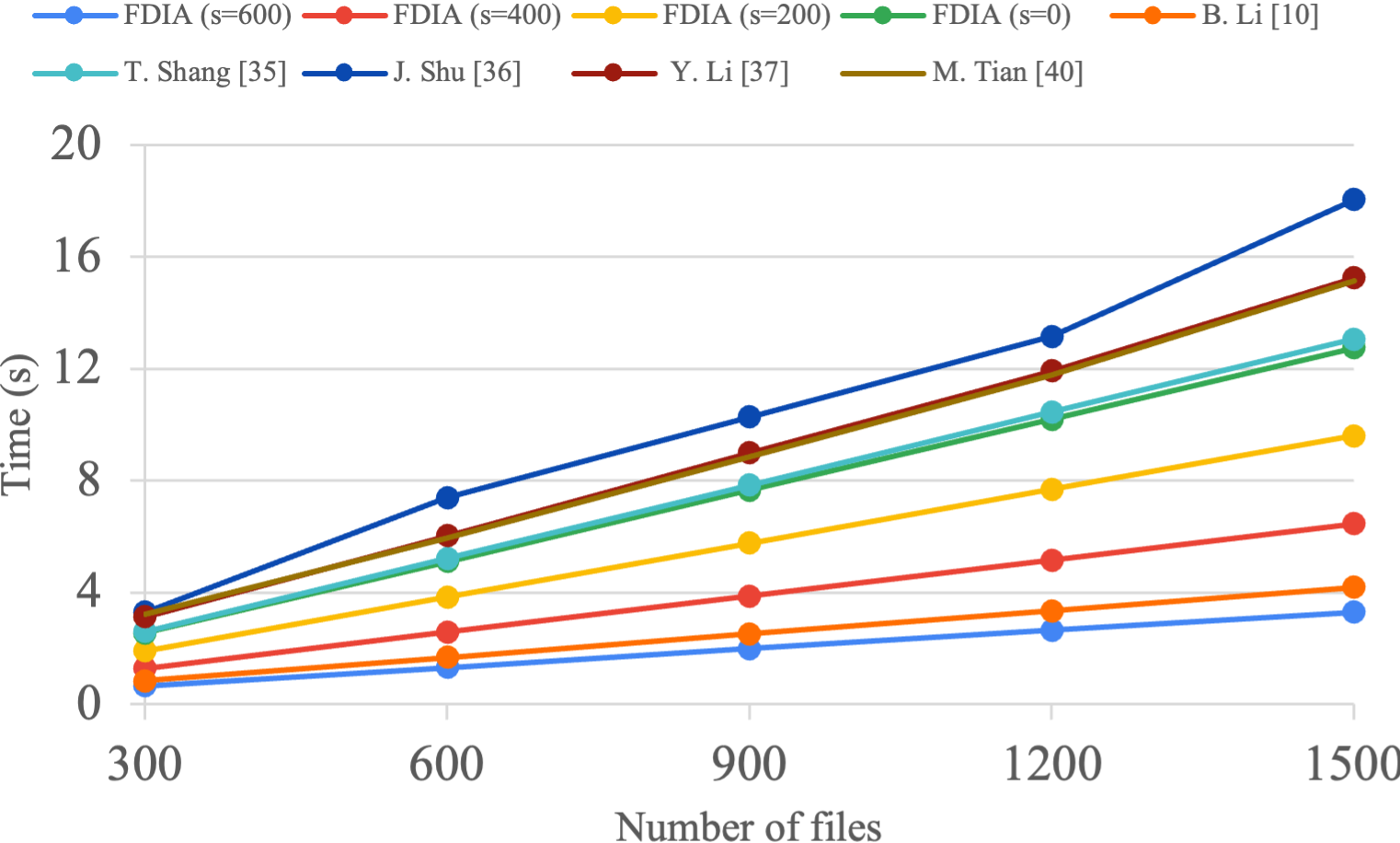}\\
		\scalebox{.5}{(e). Proof genration time ($| I | = 200$).} &
		\scalebox{.5}{(f). Proof genration time ($| I | = 400$). } &
		\scalebox{.5}{(g). Proof genration time ($| I | = 600$).} &
		\scalebox{.5}{(h). Proof genration time ($| I | = 800$).}\\
		\\
			\end{tabular}
	\caption{\footnotesize The computational overhead on the auditor and auditing ES for generating challenges and integrity proofs, respectively.}
	\label{keyCharts}
\end{figure*}
\begin{figure}
	\begin{tabular}{cc}		
		\includegraphics[width=1.66in]{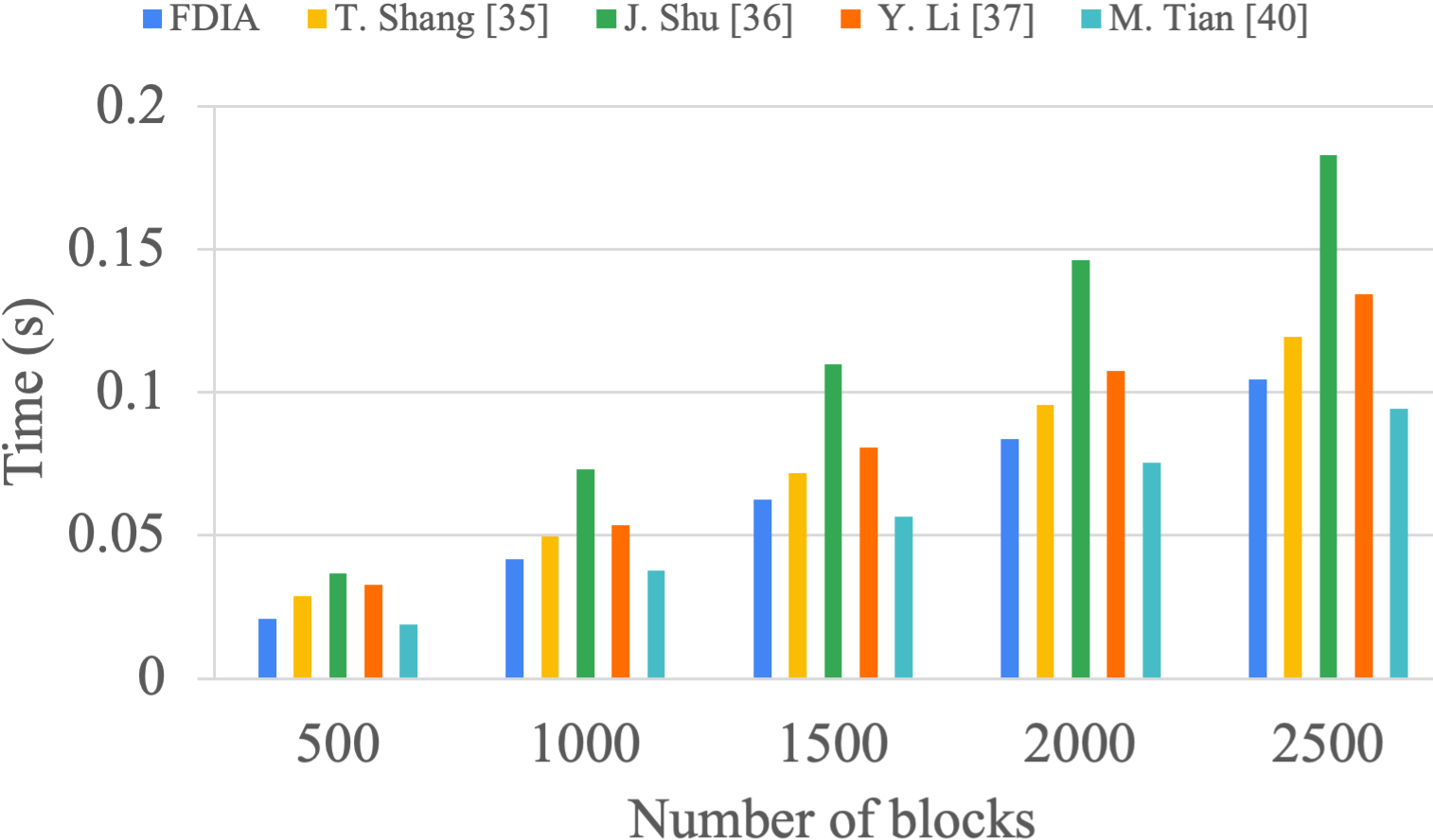}&
		\includegraphics[width=1.66in]{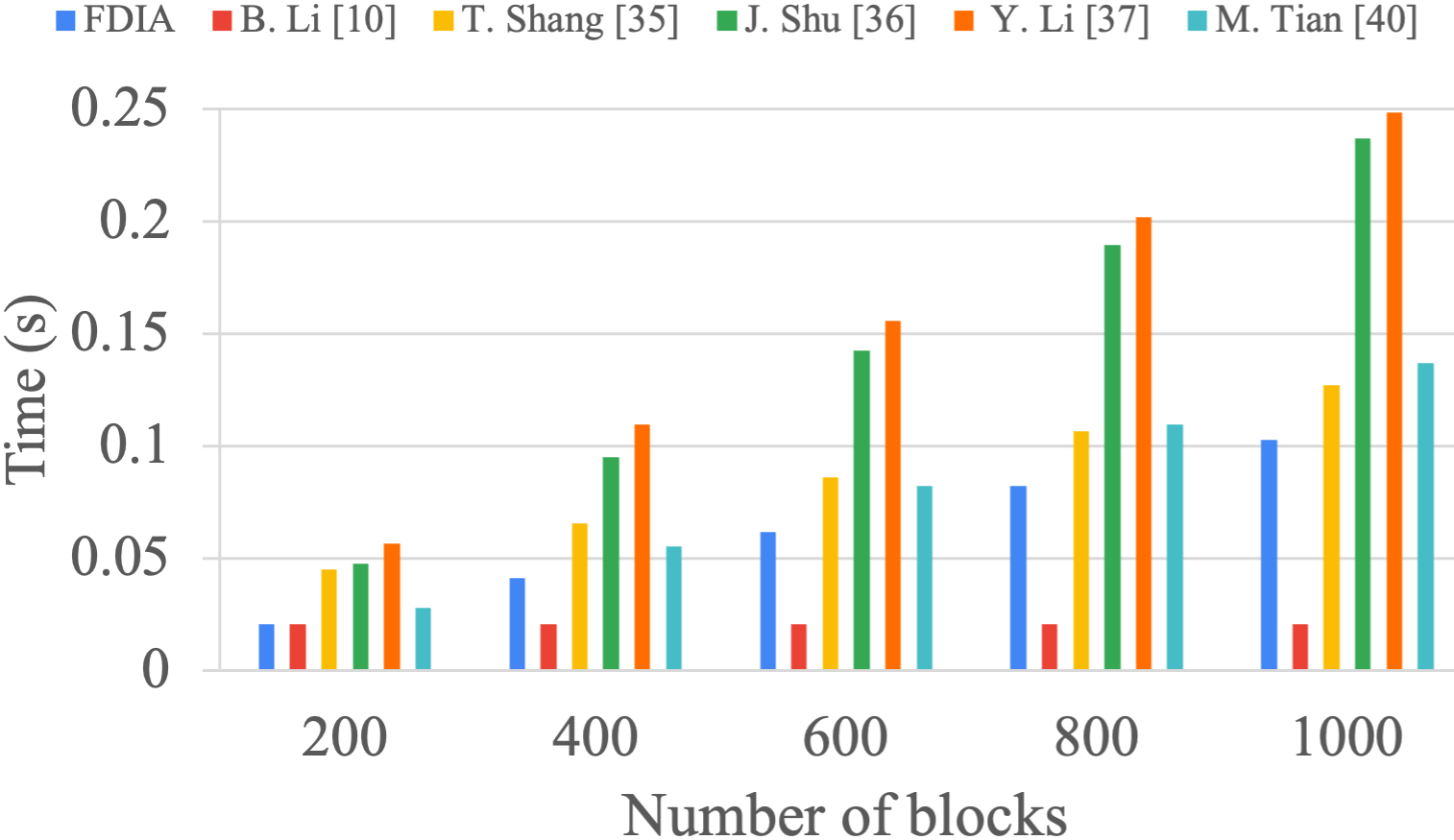}
		\\
		\scalebox{.5}{(e). Tag generation time.} &
		\scalebox{.5}{(f). Proof check time. }
	\end{tabular}
	\caption{\footnotesize The computational overhead on the auditor and the AV for generating tags and checking integrity proofs, respectively.}
	\label{keyCharts}
\end{figure}
\begin{figure*}
	\begin{tabular}{cccc}
		\includegraphics[width=1.65in]{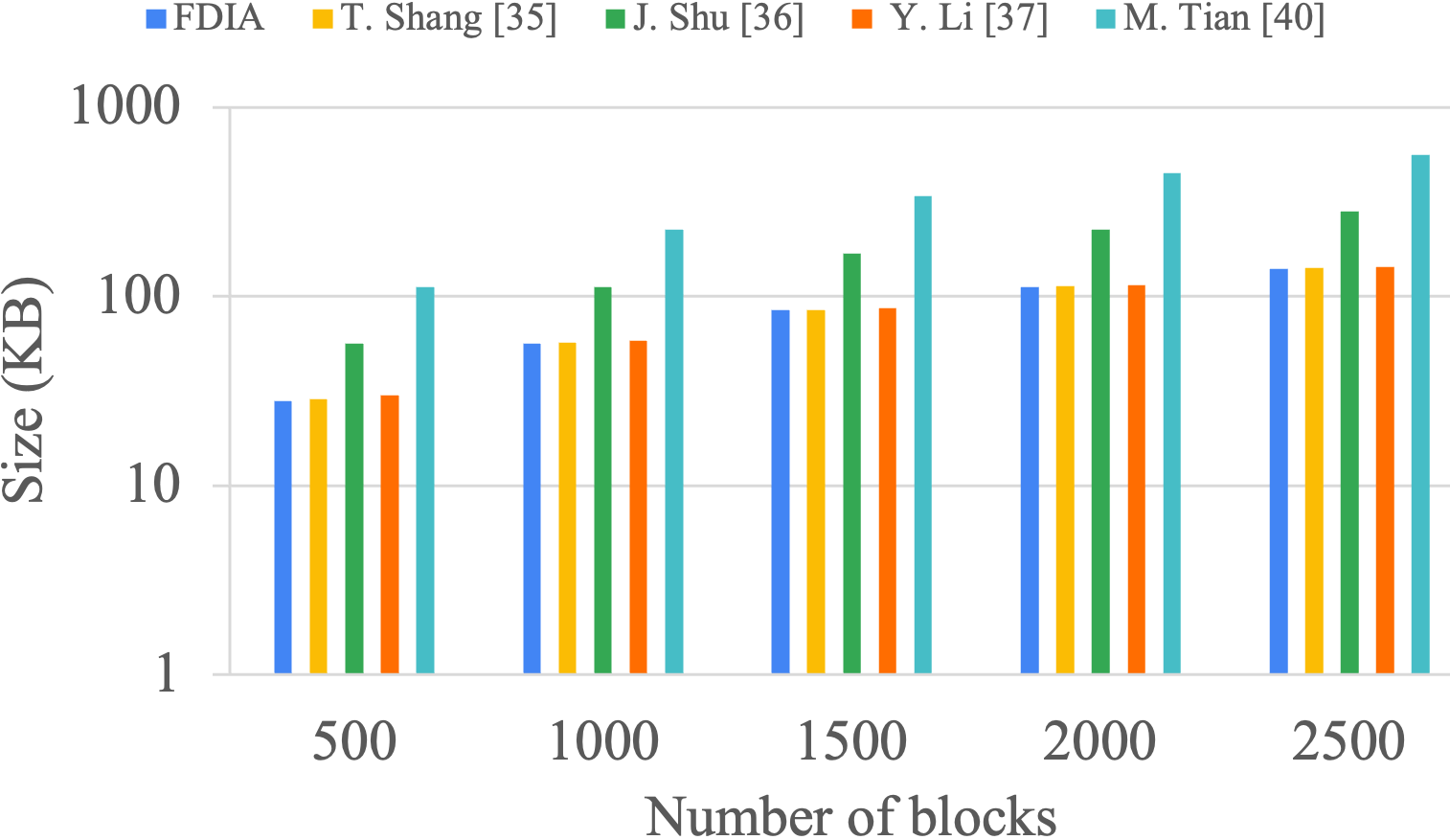}& 
		\includegraphics[width=1.65in]{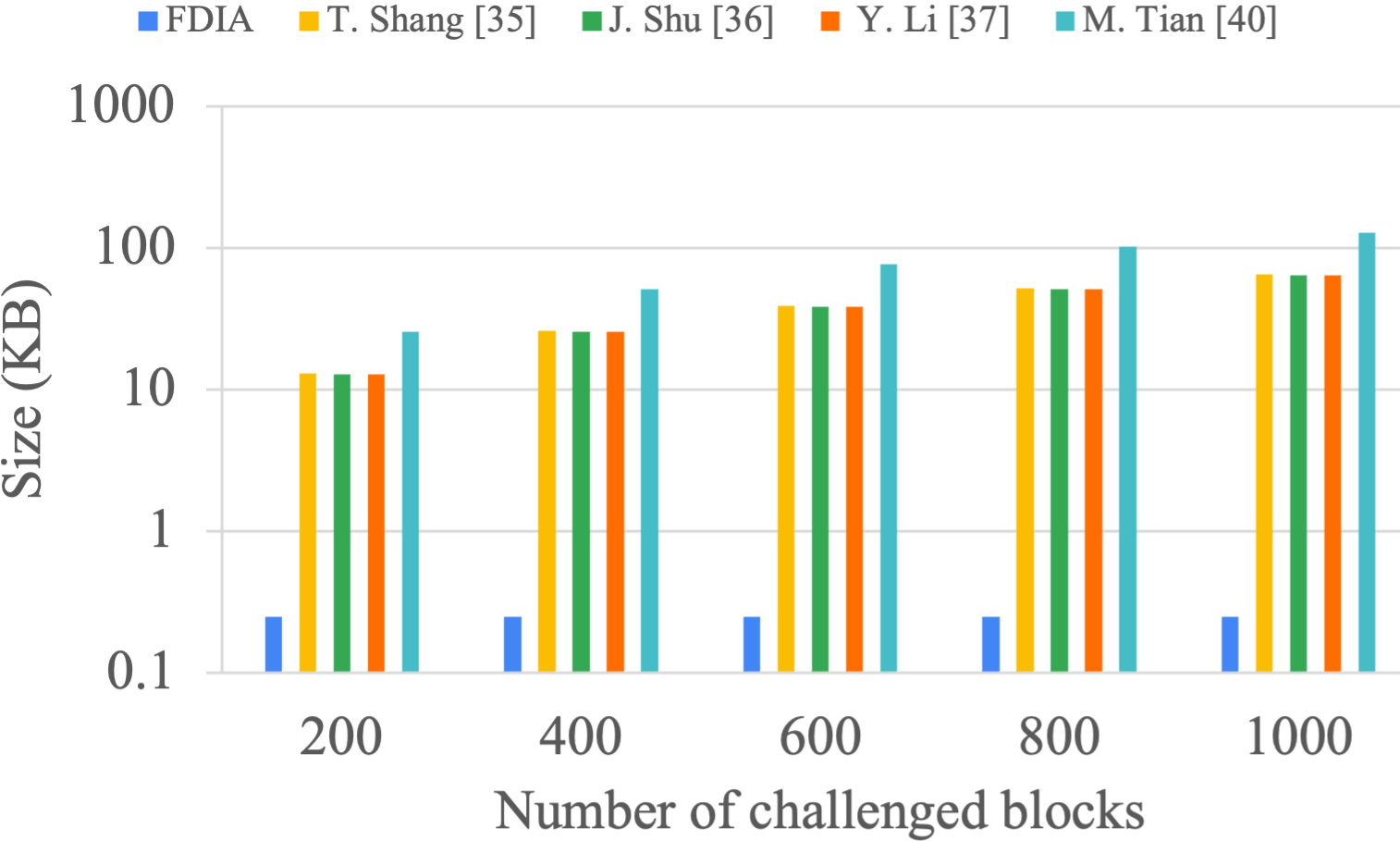}&
		\includegraphics[width=1.65in]{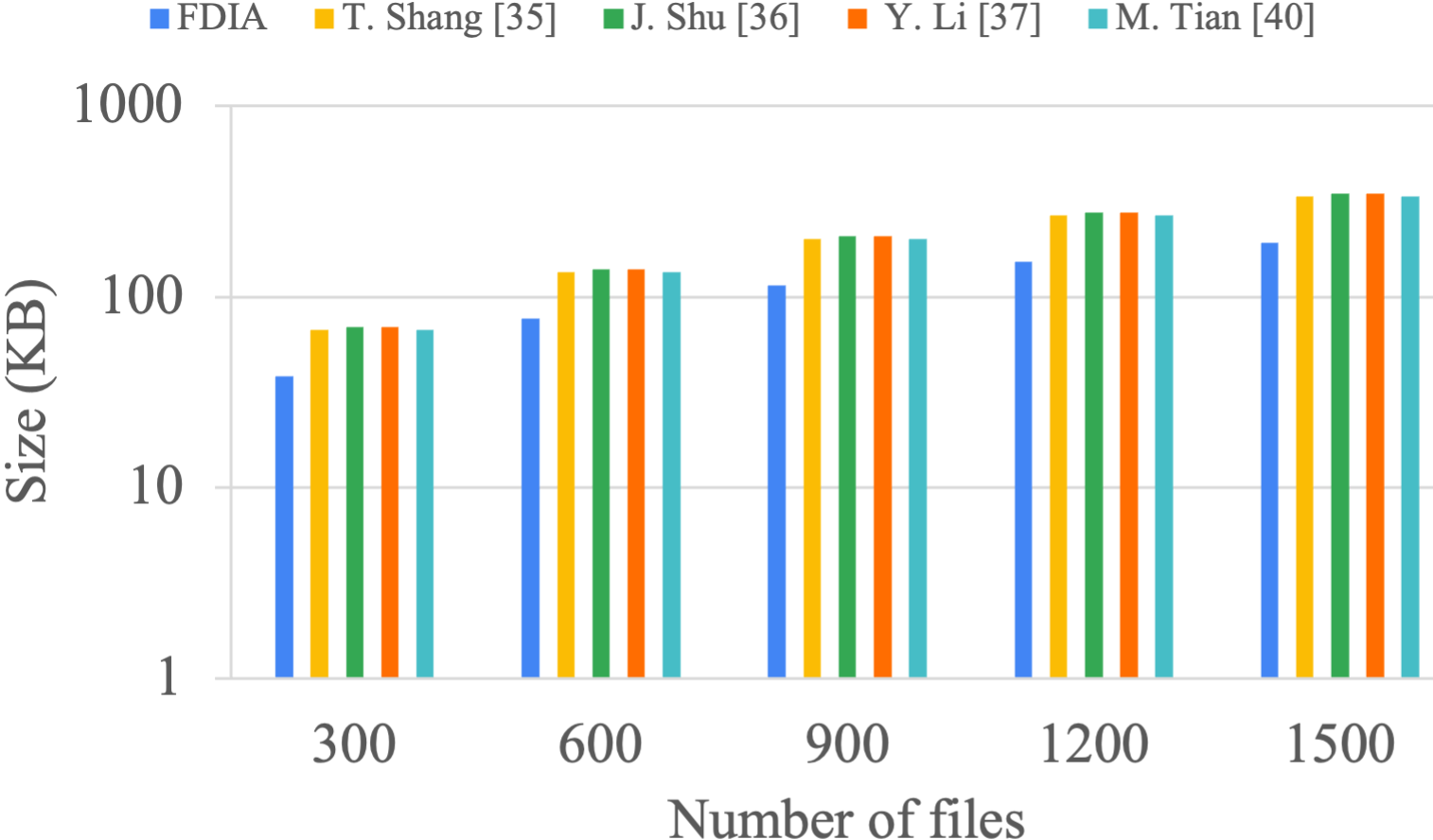}&
		\includegraphics[width=1.65in]{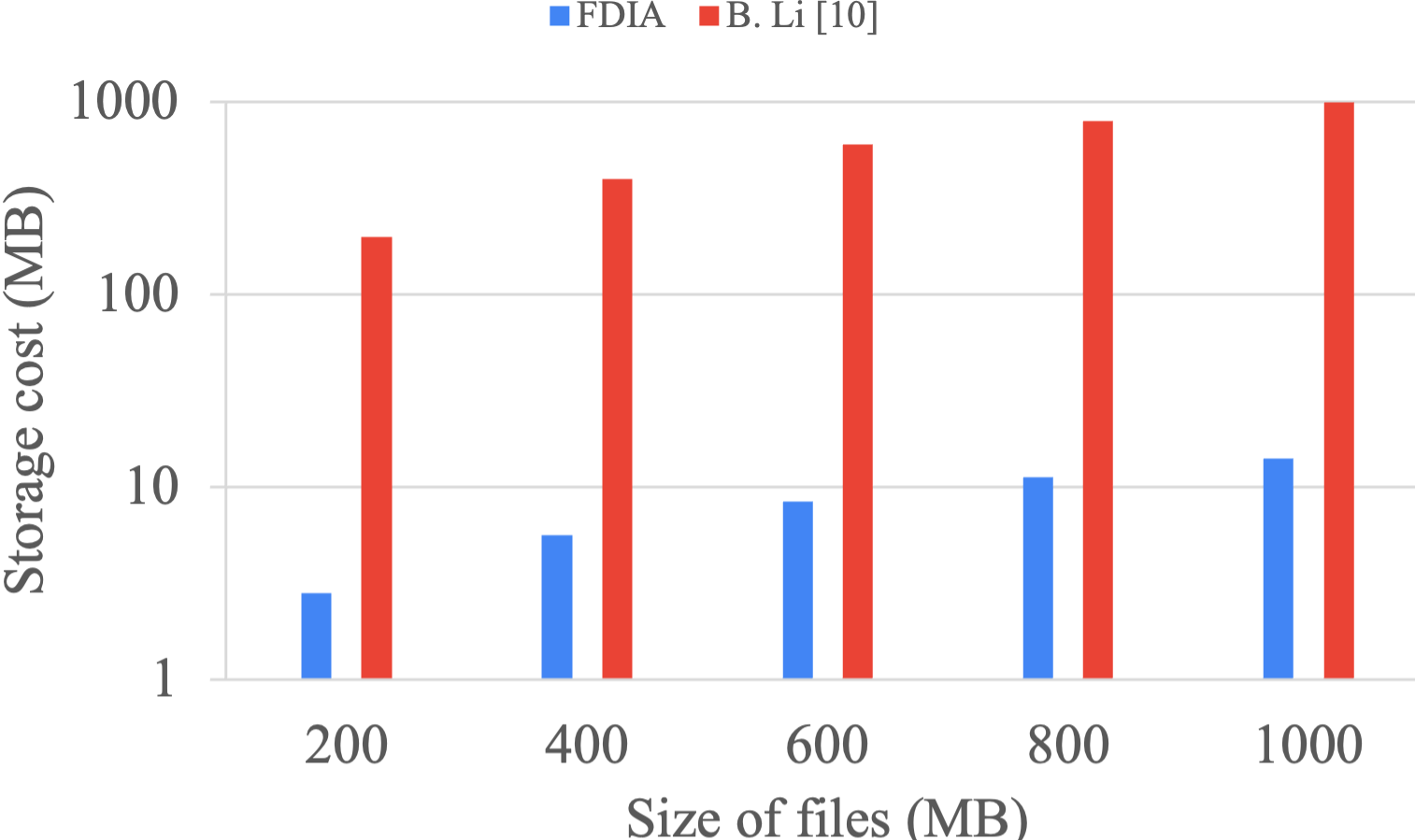}\\
		\scalebox{.5}{(a). Size of a tag.} &
		\scalebox{.5}{(b). Size of a challenge. } &
		\scalebox{.5}{(c). Size of a proof.} &
		\scalebox{.5}{(d). Storage cost for the AV. }\\
		\\
	\end{tabular}
	\caption{\footnotesize Storage overhead.}
	\label{keyCharts2}
\end{figure*}


\section{Conclusion}\label{cn}
In this paper, we presented a Fully Decentralized Integrity Auditing (FDIA) scheme, which aims to assist app vendors in verifying the integrity of their data cached in distributed edge servers. We introduced the use of edge servers as auditors and employed game theory to incentivize truthful and accurate audits. Using this approach we address concerns related to collusion between auditors and/or auditings, offering a more trustworthy auditing mechanism for app vendors and data owners. Through analysis, we demonstrated that our scheme is secure within a random oracle and thoroughly evaluated the security and performance aspects of our scheme. The results indicate that our scheme provides faster performance and proves to be a cost-effective solution compared to some mentioned approaches.\\

\textbf{Conflict of Interest}: The authors declare that they have no conflict of interest.

\bibliographystyle{IEEEtran}
\bibliography{references}

\end{document}






{\appendix[Correctness proof]
\begin{Theorem}
	Our proposed FDIA scheme is correct.
\end{Theorem}

\begin{proof}
	 We have to show that if both the auditor ES and the auditing ES are truthful, then the auditing ES can successfully pass the verification for any valid tag $t_i$ and a randomly chosen challenge. The correctness of the scheme can be elaborated as follows, where $I' = \cup_{i \in L} I'^{(j)}$:

\begin{scriptsize}
\begin{align*}
		m'& = e(\phi, \alpha) . \beta^{-\mu} = e(\prod_{j \in L}\prod_{i \in I'^{(j)}}\phi_{i} . \prod_{i \in I \setminus I'}\phi'_{i}, \alpha) .  \beta^{-\sum\limits_{j \in L}\sum\limits_{i \in I'^{(j)}} \mu_{i} + \sum\limits_{i \in I\setminus I'} \mu'_{i}} \\
		=  & e(\prod_{j \in L}\prod_{i \in I'^{(j)}}\phi_{i}, \alpha). e(\prod_{i \in I \setminus I'}\phi'_{i}, \alpha) . \beta^{-\sum\limits_{j \in L}\sum\limits_{i \in I'^{(j)}} \mu_{i}}. \beta^{\sum\limits_{i \in I\setminus I'} \mu'_{i}}\\
		 = & e(\prod_{j \in L}\prod_{i \in I'^{(j)}}\phi_{i}, \alpha). e(\prod_{i \in I \setminus I'}\phi'_{i}, \alpha) . (e(H_{1}(id), g^z)^{\lambda})^{-\sum\limits_{j \in L}\sum\limits_{i \in I'^{(j)}} \mu_{i}}. (e(H_{1}(id), g^z)^{\lambda})^{-\sum\limits_{i \in I\setminus I'} \mu'_{i}}\\
		= & e(\prod_{j \in L}\prod_{i \in I'^{(j)}}t_{i}^{f_{k^{(j)}_{PRF}}(i)}, \alpha). e(\prod_{i \in I \setminus I'}t_{i}^{f_{k_{PRF}}(i)}, \alpha) . e(s, g^{\lambda})^{-\sum\limits_{j \in L}\sum\limits_{i \in I'^{(j)}}f_{k^{(j)}_{PRF}}(i) f_i}\\
		.& e(s, g^\lambda)^{-\sum\limits_{i \in I\setminus I'} f_{k_{PRF}}(i) f_i}\\
= &\prod_{j \in L}\prod_{i \in I'^{(j)}} e(t_{i}^{f_{k^{(j)}_{PRF}}(i)}, \alpha). \prod_{i \in I \setminus I'} e(t_{i}^{f_{k_{PRF}}(i)}, \alpha) \\
.& \prod_{i \in I'?} e(s, g^{\lambda})^{-f_{k^{i}_{PRF}}(i) f_i}. \prod_{i \in I \setminus I'} e(s, g^\lambda)^{-f_{k_{PRF}}(i) f_i}\\
		= &\prod_{j \in L}\prod_{i \in I'^{(j)}} e(t_{i}s^{-f_i}, g^{\lambda})^{f_{k^{(j)}_{PRF}}(i)}. \prod_{i \in I \setminus I'} e(t_{i}s^{-f_i}, g^{\lambda})^{f_{k_{PRF}}(i)}\\
		= & \prod_{j \in L}\prod_{i \in I'^{(j)}} e(H_2(name_F||h"||i)^{f_{k_{PRF}^{(j)}}(i)} , h'^{\lambda}). \prod_{i \in I \setminus I'} e(H_2(name_F||h"||i)^{f_{k_{PRF}}(i)} , h'^{\lambda})	
	\end{align*}
\end{scriptsize}
\end{proof}

}